\documentclass[aps,prresearch,reprint,superscriptaddress,longbibliography]{revtex4-2}

\usepackage{amsmath,amssymb,bm}
\usepackage{graphicx}
\usepackage[colorlinks=true, linkcolor=blue, urlcolor=blue,citecolor=blue]{hyperref}
\usepackage{physics}

\begin{document}

\title{Real-Space Theory of Nonequilibrium Orbital Magnetization in Inhomogeneous Solids}

    \author{João M. Alendouro Pinho}
    \thanks{These authors contributed equally to this work.}
    \affiliation{Centro de Física das Universidades do Minho e Porto Departamento de Física e Astronomia,
Faculdade de Ciências, Universidade do Porto, 4169-007 Porto, Portugal}
    \author{Henrique P. Veiga}
    \thanks{These authors contributed equally to this work.}
    \affiliation{Centro de Física das Universidades do Minho e Porto Departamento de Física e Astronomia,
Faculdade de Ciências, Universidade do Porto, 4169-007 Porto, Portugal}
    \affiliation{School of Physics, Engineering and Technology and York Centre for Quantum Technologies, University of York, YO10 5DD, York, United Kingdom}
    \author{Alessandro Veneri}
    \affiliation{Dipartimento di Ingegneria dell’Informazione, Elettronica e Telecomunicazioni, Università La Sapienza, Via Eudossiana, 18, 00184 Roma, Italy}
    \affiliation{School of Physics, Engineering and Technology and York Centre for Quantum Technologies, University of York, YO10 5DD, York, United Kingdom}
    \author{Daniel Passos}
    \affiliation{Centro de Física de Materiales, Universidad del País Vasco, 20018 San Sebastián, Spain}
    \author{Ivo Souza}
    \affiliation{Centro de Física de Materiales, Universidad del País Vasco, 20018 San Sebastián, Spain}
    \affiliation{Ikerbasque Foundation, 48013 Bilbao, Spain}
    \author{Bruno Amorim}
    \affiliation{Centro de Física das Universidades do Minho e Porto Departamento de Física e Astronomia,
Faculdade de Ciências, Universidade do Porto, 4169-007 Porto, Portugal}
    \author{João M. Viana Parente Lopes}
    \affiliation{Centro de Física das Universidades do Minho e Porto Departamento de Física e Astronomia,
Faculdade de Ciências, Universidade do Porto, 4169-007 Porto, Portugal}
    \author{Aires Ferreira}
    \email{aires.ferreira@york.ac.uk}
    \affiliation{School of Physics, Engineering and Technology and York Centre for Quantum Technologies, University of York, YO10 5DD, York, United Kingdom}
\begin{abstract}
We put forward a real-space linear-response theory of orbital magnetization induced by a dc electric field in bulk materials lacking translation invariance. Starting from the current response to a slowly varying electromagnetic field, we show that the linear-in-wavevector  conductivity  tensor of non-interacting electrons can be cast entirely in terms of single-particle Green's functions and the velocity and position operators. This provides a gauge-invariant real-space formulation of the long-wavelength non-equilibrium orbital magnetization of both perfect crystals and disordered systems. In the translationally invariant limit, the formalism recovers the established Bloch picture of orbital magneto-electric effects. We implement the formalism using large-scale Chebyshev expansions  and illustrate its capabilities on a Haldane model in the presence of spatially uncorrelated disorder. 
\end{abstract}

\maketitle

\section{Introduction}

Electrical generation and control of magnetization is a central theme in magnetoelectrics, spintronics and, more recently, orbitronics \cite{Fiebig_2005,Ganichev_2003,RevModPhys.91.035004,Cullen_2026}. In inversion-asymmetric conductors, an applied electric field can generate a nonequilibrium orbital magnetization (the orbital analog of the inverse spin-galvanic effect  \cite{Ivchenko_1978,Ivchenko_1989,Aronov_1989,Edelstein_1990}). Early theoretical work established this phenomenon in chiral crystals \cite{yoda_current-induced_2015-1,yoda_orbital_2018,zhongGyrotropicMagneticEffect2016} and subsequent studies predicted large orbital responses in noncentrosymmetric antiferromagnets, polar materials, oxide interfaces and moiré systems \cite{Salemi_19,Hara_20,He_20,johansson_spin_2021}. The growing theoretical interest has been accompanied by experimental progress: current-driven switching of orbital magnetic states has been observed in moir\'e graphene \cite{sharpe_emergent_2019,serlin_intrinsic_2020}, while signatures of orbital-charge interconversion have been detected in interfaces hosting two-dimensional electron gases  \cite{el_hamdi_observation_2023,gao_nonlocal_2025}. These developments have established current-induced orbital magnetism as a promising route for electrical manipulation of magnetic states without relying exclusively on spin angular momentum.

Despite its apparently simple physical meaning, the orbital magnetization of a solid  is a subtle bulk observable. For a finite system with open boundaries one may formally write the  orbital moment   as the expectation value of the operator $-(e/2)\, \mathbf r\times\mathbf v  $, but the position operator is unbounded and ill defined in a periodic crystal, making the direct thermodynamic limit ambiguous. The modern theory of orbital magnetization resolved this problem by recasting the observable in terms of gauge-invariant bulk quantities, revealing both local-circulation and itinerant contributions \cite{Thonhauser_05_ModernT,ceresoli_orbital_2006,xiao_adiabatically_2021,shi_quantum_2007,resta_electrical_2010}. A real-space formulation valid for  inhomogeneous systems was   recently developed \cite{Vidarte_26}.

The nonequilibrium problem is yet more delicate. In a current-carrying state, a naive linear-response real-space evaluation of the expectation value $\langle\mathbf r\times\mathbf v\rangle$  is origin dependent, since a shift of the coordinate origin produces a term proportional to the transport current. A meaningful quantum treatment of nonequilibrium orbital phenomena must therefore separate circulating currents from homogeneous transport currents. Two complementary routes have been pursued to achieve this: semiclassical  transport and wave-packet formulations express the magneto-electric response in terms of Bloch-state orbital moments, band geometry, and field-induced changes to the electronic distribution function \cite{yoda_current-induced_2015-1,yoda_orbital_2018,zhongGyrotropicMagneticEffect2016,He_20,xiao_adiabatically_2021,Cullen_2026}, while fully quantum-mechanical approaches based upon linear-response Kubo theory extract the orbital magnetization from the  linear-in-wavevector spatially dispersive conductivity \cite{malashevichBandTheorySpatial2010,shinadaQuantumTheoryIntrinsic2023,Ocana_23}. These theories are constructed in momentum space and therefore rely on translational invariance.

On the other hand, comparatively little is known about the effect of disorder and lattice defects, particularly concerning the prevalence and efficiency of extrinsic mechanisms of charge-to-orbital conversion. Some studies have attempted to model disorder effects by means of a phenomenological impurity self-energy (i.e. a spectral broadening), which preserves the basic features of the intrinsic  magneto-electric response and does not give access to the underlying orbital scattering mechanisms that may be active in a material. A few works have treated disorder microscopically, though, establishing the relative contributions from quantum side-jump processes, skew scattering and related extrinsic mechanisms to current-induced orbital magnetization  \cite{rouKineticOrbitalMoments2017,sahin_pancharatnam-berry_2018-1,Liu_24,Veneri_25,sarkar_extrinsic_2026,cullen_theory_2026}. These approaches, however, treat disorder perturbatively and become extremely involved for multi-band models and impurity potentials of realistic complexity \cite{Veneri_25} (see also Ref. \cite{Offidani_18} in the related context of the spin and anomalous Hall effects). Despite their success in identifying key extrinsic mechanisms for electrical generation of orbital Hall currents and orbital magnetization due to scalar random potentials, Bloch-space perturbative methods are not designed for addressing extended defects and local lattice distortions, spatial inhomogeneities or coherent multiple-scattering processes beyond the weak-disorder (Born) limit. Moreover, strong short-range scatterers, such as point defects and resonant impurities,   dictate the transport properties of important material systems \cite{Rutter_07,Yazyev_07,Alloul_09,Ferreira_11,Ugeda_11,Hong_15,Jiang_18,Barja_19,Joucken_21,Wang_23}, but their impact on orbital transport is challenging to study with existing techniques \cite{Veneri_25}. It would thus be desirable to have an alternative approach that is free of such limitations.

In this paper, we develop a fully quantum-mechanical  formulation of dc electric-field induced orbital magnetization that does not require translational symmetry and  is amenable to efficient numerically exact simulations. Our starting point is the spatially dispersive electromagnetic response [see Eq.\,\,(\ref{eq:current}) below], which provides a rigorous foundation to understand orbital magneto-electric effects in inhomogeneous solids. Our central result is a gauge-invariant Kubo-like formula for the magneto-electric tensor written in terms of single-particle Green's functions and velocity and position operators. An alternative but equivalent construction based solely on Green's functions and velocity operators is also presented. 

The remainder of the paper is organized as follows. Section \ref{sec:II-basic_framework} sets up the general problem and summarizes the current understanding. In Sec. \ref{sec:III-generalised_current_vertex}, we derive a general (basis-independent) expression for the electromagnetic response kernel, which will form the basis of our investigation. Further, we carefully examine the dc limit of the linear-in-wavevector conductivity and obtain the key formal results of this work in Sec. \ref{sec:IV-dc_limit}. Section \ref{sec:V-Onsager_decomposition} utilizes a symmetry-based St\v{r}eda decomposition to isolate the time-even and time-odd Onsager sectors of  the magneto-electric response. In Sec. \ref{sec:VI-Results}, we employ the new real-space formalism to investigate the non-equilibrium orbital magnetization of the anisotropic Haldane model with spatially uncorrelated disorder. In particular, we demonstrate that the presence of a random scattering potential strongly reshapes the magnetoelectric response, even for weak-disorder limit. Finally, Sec. \ref{sec:VII-Conclusions} summarizes our conclusions and prospects for future work. The technical proofs are given in the Appendices.

\section{Current-induced orbital magnetization}
\label{sec:II-basic_framework}

In the following, we outline the linear-response formalism used to compute the dc electric-field induced magnetization  of disordered non-interacting systems. We begin with a brief theoretical background. The central idea is that non-equilibrium magnetization ensues from the transverse, circulating component of the current response to an external perturbation. In this work, we take as the perturbation a monochromatic electric field with a slow temporal modulation $\omega$ and a long spatial modulation wavevector ${\bf q}$. In this long-wavelength regime, the linear-response current admits the expansion 
\begin{equation} \label{eq:current}
J_a(\mathbf{q},\omega)=\sigma_{ab}(\omega)E_b(\mathbf{q},\omega)+\sigma_{ab,c}(\omega)q_{c}E_b(\mathbf{q},\omega)+O\left(q^{2}\right),
\end{equation}
with the usual summation convention over repeated spatial indices in effect. Here, $\sigma_{ab}$ describes the spatially uniform electric-dipole component of the induced charge current distribution. Moreover, the linear-in-$\mathbf q$  term in Eq.~(\ref{eq:current})   is the leading spatially dispersive correction and captures magnetic-dipole and electric-quadrupole contributions to the current response. At finite frequency, the rank-3 conductivity tensor $\sigma_{ab,c}$ describes spatially dispersive optical phenomena such as optical activity \cite{landau_electrodynamics_1984} and gyrotropic birefringence \cite{Hornreich_68}.

The magnetization induced by a homogeneous dc electric field can be extracted from the leading spatially-dispersive term in Eq.~(\ref{eq:current}) \cite{Levitov_85,Hornreich_68}. Evaluation of the circulating part of $\mathbf{J}(\mathbf{q},\omega)$ in the $\omega\rightarrow 0$ limit leads to the linear-response expression for the dc non-equilibrium magnetization induced by an applied dc electric field:
\begin{equation} \label{eq:magnetoelectric_tensor}
\delta M_{b}  = K_{ab}\, E_{a},
\end{equation}
where $K$ is the linear magneto-electric response tensor. This quantity can be separated into time-odd and time-even components, $K=\tilde\alpha + \beta$. The two contributions are obtained directly from the corresponding
Onsager sectors of the linear-in-$\mathbf q$ conductivity 
as ~\cite{Hornreich_68,malashevich_theory_2010,Ocana_23,shinadaQuantumTheoryIntrinsic2023}
\begin{subequations}
\label{eq:magnetos}
\begin{align}
\tilde{\alpha}_{ab} & =\frac{1}{3i}\epsilon_{cdb}\sigma^{\mathrm{S}}_{ac,d}(\omega\rightarrow0), \label{odd-magnetoelectric}\\
\beta_{ab} & =\frac{1}{4i}\epsilon_{cdb}\left[\sigma^{\mathrm{A}}_{cd,a}(\omega\rightarrow0)-2\sigma^{\mathrm{A}}_{ac,d}(\omega\rightarrow0)\right], \label{even-magnetoelectric}
\end{align}
\end{subequations}
where $\sigma^{\mathrm{S(A)}}_{ab,c}$ denotes the symmetric (antisymmetric) part with respect to the interchange $a\leftrightarrow b$.

Both of those sectors require broken inversion symmetry, whilst only the symmetric (time-odd)  part requires broken time-reversal symmetry as well. Moreover, the antisymmetric (time-even) part vanishes if both inversion and time-reversal are broken but the combined $\mathcal{PT}$ symmetry is preserved. 

We further note that the tensor $\beta$ contains nine independent quantities, whilst $\tilde\alpha$ has only eight. In fact, $\tilde\alpha$ corresponds to the traceless part of the time-odd linear magnetoelectric tensor
$\alpha_{ab} = \tilde\alpha_{ab} + (\theta e^2/2\pi h)\delta_{ab}$. The trace piece $\theta$, known as the axion, cannot be recovered from the spatially-dispersive conductivity and must be obtained by other means \cite{malashevich_theory_2010,essin_orbital_2010}. The time-even magnetoelectric effect described by $\beta$ is dissipative and only occurs in conductors, whilst the time-odd effect described by $\alpha$ is reactive and occurs in insulators as well. In microscopic theories, the tensors $\alpha$ and $\beta$ generally comprise both orbital and spin contributions; for the spinless model considered in this work, the spin part is absent.

In two recent works~\cite{Ocana_23,shinadaQuantumTheoryIntrinsic2023}, the spatially-dispersive conductivity $\sigma_{ab,c}$ was evaluated in the independent-particle approximation for pristine conducting crystals, and the magneto-electric response was recovered via \eqref{eq:magnetos}, along with the electric-quadrupole response. Previously, semiclassical formulations of those effects had been introduced \cite{gao_nonreciprocal_2019-1,lapa_semiclassical_2019,xiao_adiabatically_2021,Xiao_21}. To go beyond previous approaches, we dedicate the next sections to putting forward a real-space framework for the computation of the rank-3 conductivity tensor $\sigma_{ab,c}$, enabling the evaluation of the magnetoelectric response for generic broken translation-invariance bulk systems. 

\section{Electromagnetic coupling and generalized current vertex}
\label{sec:III-generalised_current_vertex}

The aim is to derive a basis-independent formulation for the
three-tensor conductivity $\sigma_{ab,c}$. To this end, we consider a generic
noninteracting system $\hat{H}_{0}$ coupled to an electromagnetic
perturbation $\hat{V}\left(\mathbf{\hat r},t\right)$,
\begin{equation}
\hat{H}\left(\mathbf{\hat r} , t\right)=\hat{H}_{0}\left(\mathbf{\hat r}\right)+\hat{V}\left(\mathbf{\hat r},t\right),\label{eq:H0}
\end{equation}
where $\mathbf{\hat r}$ is the position operator. The perturbation can be expanded in terms of the electromagnetic
gauge field
\begin{align}
\hat V (t)
&=
\frac{ie}{\hbar}\int d\mathbf{q}\,
v_i(\mathbf{q})\,
A^i(\mathbf{q},t)
\nonumber\\
&\quad+
\frac{-e^2}{\hbar^2}\int d\mathbf{q}\,d\mathbf{q}'\,
\Delta_{ij}(\mathbf{q},\mathbf{q}')
A^i(\mathbf{q},t)
A^j(\mathbf{q}',t)
\nonumber\\
&\quad+
O(A^3).
\label{eq:perturbation_expansion}
\end{align}
with the vector potential taken in the Weyl gauge $\partial_{t}\mathbf{A}(\mathbf{r},t)=-\mathbf{E}(\mathbf{r},t)$ and $v_{i}\left(\mathbf{q}\right)$ and $\Delta_{ij}\left(\mathbf{q},\mathbf{q}'\right)$
being the paramagnetic and diamagnetic operator densities in Fourier
space. It can be proven that, for either a tight-binding or a continuum
Hamiltonian, the paramagnetic density is written, to first order in
$\mathbf{q}$, as (see Appendix \ref{Ap:current_vertex}
for derivation)

\begin{equation}
\hat{v}_{a}(\mathbf{q})=\hat{v}_{a}+\frac{i}{2}q_{c}\{\hat{r}_{c},\hat{v}_{a}\}+\mathcal{O}(q^{2}).\label{eq:vq_expansion}
\end{equation}
where $\hat{v}_{a}=-\frac{i}{\hbar}[\hat{H}_{0},\hat{r}_{a}]$ is
the velocity operator of the unperturbed system. 
The appearance of $\{\hat{r}_{c},\hat{v}_{a}\}$ is the real-space counterpart
of the momentum derivatives entering spatially dispersive response
functions in the Bloch representation. Although in this work we are
mostly interested in the application of the formalism to tight-binding 
models, it can be proven that, to first order and without loss
of generality, Eq. \eqref{eq:vq_expansion} 
remain valid to first order in \textbf{$\mathbf{q}$ }irrespectively
of the microscopic details of the Hamiltonian \cite{mckay_charge_2024}. The 
formalism developed here is therefore valid in general and can be used to study,
for example, continuum relativistic and semi-relativistic Hamiltonians.

In principle, the induced
current at momentum $\mathbf{q}$ may couple to vector-potential components
at all momenta,

\begin{equation}
J_{a}(\omega,\mathbf{q})=\int d\mathbf{q}'\Phi_{ab}(\omega,\mathbf{q},-\mathbf{q}')A_{b}(\omega,\mathbf{q}'),
\end{equation}
with the electromagnetic response function connected to the conductivity according to 
\begin{equation}
\sigma_{ab}(\omega,\mathbf{q}) = \Phi_{ab}(\omega,\mathbf{q})/i\omega.
\label{eq:cond_phi}
\end{equation}
Specializing to systems possessing discrete translational symmetry (see below for the disordered case), momentum is
conserved modulo a reciprocal lattice vector, such that  

\begin{equation}
\Phi_{ab}(\omega,\mathbf{q},-\mathbf{q}')=\sum_{\mathbf{G}}\Phi^{(\mathbf{G})}_{ab}(\omega,\mathbf{q})\,\delta\left(\mathbf{q}-\mathbf{q}'-\mathbf{G}\right),
\end{equation}
and the current becomes

\begin{equation}
J_{a}(\omega,\mathbf{q})=\sum_{\mathbf{G}}\Phi^{(\mathbf{G})}_{ab}(\omega,\mathbf{q})A_{b}(\omega,\mathbf{q}-\mathbf{G}).\label{Finite G Current}
\end{equation}

The terms with $\mathbf{G}\neq0$ correspond to Bragg-reflected processes, describing
charge current modulations on the scale of the unit cell. While
such contributions may be relevant for the study of microscopic currents,
the main focus of the present work is the macroscopic response. Therefore,
we only consider processes with $\mathbf{G}=0$, simplifying Eq.$\,$\eqref{Finite G Current}
to 
\begin{equation}
J_{a}(\omega,\mathbf{q})
=
\Phi_{ab}(\omega,\mathbf{q})
A_{b}(\omega,\mathbf{q}),
\label{eq:current_q}
\end{equation}
where 
\begin{equation}
\Phi_{ab}(\omega,\mathbf{q})
\equiv
\Phi_{ab}(\omega,\mathbf{q},-\mathbf{q}).
\end{equation}
For a single disorder realization translational symmetry is generally
absent and Eq.~(\ref{eq:current_q}) no longer holds. However, if
the disorder ensemble is statistically homogeneous, translational
invariance is restored after disorder averaging. In this case, the
disorder-averaged response kernel is again diagonal in momentum space,
allowing the same macroscopic description of Eq.~(\ref{eq:current_q})
to be employed. 
Performing a standard Kubo calculation, we may formulate this coefficient in a basis-independent manner as (see Appendix \ref{Ap:Linear_Response_Theory} for details)

\begin{equation}
\begin{aligned}
\Phi_{ab}(\omega,\mathbf{q})
&=
\frac{ie^2}{2 \pi \hbar}
\int d\varepsilon\,
f(\varepsilon)\,
\mathrm{Tr}\Big[
\Delta \hat G(\varepsilon)
\hat v_{a}(\mathbf{q})
\\
&\qquad\times
\hat G^{R}(\varepsilon+\omega)
\hat v_{b}(-\mathbf{q})
\\
&\quad+
\Delta \hat G(\varepsilon)
\hat v_{b}(-\mathbf{q})
\hat G^{A}(\varepsilon-\omega)
\hat v_{a}(\mathbf{q})
\Big].
\end{aligned}
\label{eq:Phi_general}
\end{equation}

In the above, $G^{R/A}$ denote retarded/advanced Green's functions of the unperturbed system and $\Delta G = G ^R - G ^ A $. Equation \eqref{eq:Phi_general} encodes the linear response to a dispersive field with arbitrary wave-vector, written in a basis independent formulation. When applied to a translationally-invariant system, Eq. \eqref{eq:Phi_general} reduces to the coefficient previously derived for Bloch electrons in Ref. \cite{shinadaOrbitalOpticalActivity2024} (see Appendix \ref{Ap:Crystalline_Bloch_Limit} for proof). To obtain the rank-3 conductivity present in Eqs. \eqref{odd-magnetoelectric} and \eqref{even-magnetoelectric}, we expand the Kubo kernel to linear order in $\mathbf{q}$,
\begin{equation}
\Phi_{ab}(\omega,\mathbf{q})=\Phi_{ab}(\omega)+\Phi_{ab,c}(\omega)q_{c}+O(q^{2}),
\end{equation}
where 
\begin{equation}
\Phi_{ab,c}(\omega)
=
\left.
\frac{
\partial\Phi_{ab}(\omega,\mathbf{q})
}{
\partial q_{c}
}
\right|_{\mathbf{q}=0}
\end{equation}
constitutes the central object of the present work. We note that the zeroth-order term corresponds to the conventional response to a spatially uniform electric field and will not be considered further. 
Substituting Eq. \eqref{eq:vq_expansion} into Eq. \eqref{eq:Phi_general} we get
\begin{align}
\Phi_{ab,c}(\omega)
&=
\frac{ie^2}{2\hbar}
\int d\varepsilon\,
f(\varepsilon)\,
\mathrm{Tr}\Big[
\Delta \hat G(\varepsilon)
\{\hat r_{c},\hat v_{a}\}
\nonumber\\
&\qquad\times
\hat G^{R}(\varepsilon+\omega)
\hat v_{b}
\nonumber\\
&\quad-
\Delta \hat G(\varepsilon)
\hat v_{a}
\hat G^{R}(\varepsilon+\omega)
\{\hat r_{c},\hat v_{b}\}
\nonumber\\
&\quad-
\Delta \hat G(\varepsilon)
\{\hat r_{c},\hat v_{b}\}
\hat G^{A}(\varepsilon-\omega)
\hat v_{a}
\nonumber\\
&\quad+
\Delta \hat G(\varepsilon)
\hat v_{b}
\hat G^{A}(\varepsilon-\omega)
\{\hat r_{c},\hat v_{a}\}
\Big],
\label{eq:Phi_ijc_trace}
\end{align}
corresponding to the fundamental expression from which all subsequent
results are derived. 
We will dedicate the next section to extracting the linear-in-wavector conductivity in the dc limit. This will be key for the extraction
of the real-space magneto-electric tensor.

\section{dc limit} 
\label{sec:IV-dc_limit}

In the study of the electromagnetic response, the order of limits is crucial. Sending $\omega\rightarrow0$ first and retaining a finite $\mathbf{q}$ isolates the current response to a static vector potential, which equates to calculating the response to a static magnetic field. Conversely, sending $\mathbf{q}$ to zero first describes, among other things, the response to a uniform electric field. It is  instructive to examine the response
to a static but spatially non-uniform gauge potential before taking $\omega\rightarrow0$. Formally, this
corresponds to expanding Eq.~(\ref{eq:Phi_ijc_trace}) around $\omega=0$,
\begin{equation}
\Phi_{ab,c}(\omega)=\Phi_{ab,c}(0)+\omega\,\partial_{\omega}\Phi_{ab,c}(0)+O(\omega^{2}).\label{eq:phi_omega_expansion}
\end{equation}
Since the conductivity is related to the response kernel through Eq. \eqref{eq:cond_phi}, a finite value of $\Phi_{ab,c}(0)$ would generate a singular contribution
in the dc limit. Gauge invariance implies $\Phi_{ab,c}(0)=0$ for an applied longitudinal electric field, $\mathbf{q}\parallel \mathbf{E}$ \cite{rammer2004quantum,Rostami_2021}.
For transverse electric fields, additional constraints are necessary to determine
the vanishing of $\Phi_{ab,c}(0)$. For Bloch electrons, it has been shown that the static
contribution vanishes as a consequence of topological constraints,
leading to a no-go theorem for the chiral magnetic effect in equilibrium
\citep{zhongGyrotropicMagneticEffect2016}. For two-dimensional systems, we provide an alternative proof of the vanishing of the static contribution in Appendix \ref{Ap:Static_Contribution} that is valid in general, i.e. for both longitudinal and transverse electric fields. Therefore, in the present work, we can safely disregard this term. With this said, the dc linear-in-wavevector conductivity can be written as (see Appendix \ref{Ap:dc_limit_derivation})
\begin{align}
\sigma_{ab,c}
&=
\frac{ie^2}{2\pi\hbar}
\int d\varepsilon\,
f(\varepsilon)\,
\mathrm{Re}\,
\mathrm{Tr}\Big[
\Delta \hat G\left(\varepsilon\right)
\{\hat r_{c},\hat v_{a}\}
\nonumber\\
&\qquad\times
\partial_{\varepsilon}\hat G^{R}(\varepsilon)
\hat v_{b}
\nonumber\\
&\quad-
\Delta \hat G\left(\varepsilon\right)
\hat v_{a}
\partial_{\varepsilon}\hat G^{R}(\varepsilon)
\{\hat r_{c},\hat v_{b}\}
\Big].
\label{eq:dc_conductivity_linear_q}
\end{align}

Equation~\eqref{eq:dc_conductivity_linear_q} is the central real-space result of this work. It requires no assumption of translational symmetry and can therefore be evaluated directly for disordered or otherwise inhomogeneous tight-binding systems.
Although Eq.~\eqref{eq:dc_conductivity_linear_q} contains the position operator explicitly, the physical response is independent of the arbitrary choice of coordinate origin. To see this, expand the anticommutators and use the cyclicity of the trace to write
\begin{align}
\sigma_{ab,c}
=&\frac{ie^2}{2\pi\hbar}
\int d\varepsilon\,f(\varepsilon)
\operatorname{ReTr}\Big[
[\Delta\hat G,\hat r_c]
\hat v_a\partial_\varepsilon\hat G^R\hat v_b
\nonumber\\
&-\Delta\hat G\hat v_a
[\partial_\varepsilon\hat G^R,\hat r_c]
\hat v_b
\Big].
\label{eq:sigma_comm}
\end{align}
Clearly, an origin shift, $\hat r_c\rightarrow\hat r_c+r_{0c}$, leaves the commutators unchanged and therefore does not affect $\sigma_{ab,c}$.

Furthermore, the position operator can be eliminated completely by using
\begin{equation}
[\hat r_c,\hat G^{R/A}]
=i\hbar\hat G^{R/A}\hat v_c\hat G^{R/A}.
\label{eq:G_comm}
\end{equation}
This leads to a velocity-only formulation of the response,
\begin{align}
\sigma_{ab,c}
=&-\frac{ie^2}{2\pi}
\int d\varepsilon\,f(\varepsilon)
\operatorname{ImTr}\Big[
\big(
\hat G^R\hat v_c\hat G^R
-\hat G^A\hat v_c\hat G^A
\big) \nonumber\\
& \qquad\qquad\qquad\times\hat v_a\partial_\varepsilon\hat G^R\hat v_b \nonumber\\
&\qquad\qquad\qquad-\Delta\hat G\hat v_a
\partial_\varepsilon
(\hat G^R\hat v_c\hat G^R)
\hat v_b
\Big].
\label{eq:sigma_velocity_only}
\end{align}

This result is closely related to the modern theory of polarization
and orbital magnetization, where physically measurable quantities
are likewise expressed through commutators involving the position
operator rather than through the position operator itself, thereby
avoiding ambiguities associated with boundary conditions and the choice
of coordinate origin \citep{resta_quantum-mechanical_1998,resta_electrical_2010}. Equation \eqref{eq:sigma_velocity_only} thus poses an alternative formulation to that of Eq. \eqref{eq:dc_conductivity_linear_q}, in which the dependence on the position operator is effectively eliminated, being suitable for systems where this operator is not well defined, i.e. systems with periodic boundary conditions. The tradeoff in terms of numerical complexity is the appearance of three spectral operators, which makes \eqref{eq:sigma_velocity_only} computationally more demanding.

\section{Onsager decomposition of the orbital magnetoelectric response} \label{sec:V-Onsager_decomposition}
As a final step to extract the non-equilibrium magnetization through Eqs. \eqref{eq:magnetoelectric_tensor}, \eqref{odd-magnetoelectric} and \eqref{even-magnetoelectric}, we next separate the current response into parity sectors regarding time-reversal transformation. In linear-response theory, the Onsager reciprocity relations provide a clean way of separating these sectors in terms of symmetry classes of permutations of the three-tensor conductivity $\sigma_{ab,c}$ \cite{malashevichBandTheorySpatial2010,Ocana_23}. Specifically, the time-even $\sigma^\text{A}_{ab,c}$ and time-odd sector $\sigma^\text{S}_{ab,c}$ are given by (see Appendix \ref{Ap:Streda_Decomposition}) 

\begin{subequations}
\label{eq:onsager_sectors}
\begin{align}
\sigma^{\text{A}}_{ab,c}\left(\omega\right) & =\frac{1}{2}\left(\sigma_{ab,c}\left(\omega\right)-\sigma_{ba,c}\left(\omega\right)\right), \label{eq:onsager_even}\\ 
\sigma^{\text{S}}_{ab,c}\left(\omega\right) & =\frac{1}{2}\left(\sigma_{ab,c}\left(\omega\right)+\sigma_{ba,c}\left(\omega\right)\right). \label{eq:onsager_odd}
\end{align}

\end{subequations}

Operationally, this is the analogue of performing a generalized Středa
decomposition of the linear-response Kubo-Bastin formula as put forward in Ref. \citep{bonbien_symmetrized_2020}. When applied to the uniform 
conductivity, this decomposition has been used, despite some caveats, to distinguish between Fermi-surface
and Fermi-sea effects, commonly associated with extrinsic
and intrinsic transport \cite{Joao_2025}, respectively. For a dispersive response, however, this isolation of Fermi-surface and Fermi-sea contributions is generically lost \citep{bonbien_symmetrized_2020,pozo_ocana_multipole_2023}. Nevertheless, it provides insight into the symmetries of the model, as this two sectors are still assured by Onsager relations to behave differently under symmetry operations involving time-reversal. 

Applying Eqs. \eqref{eq:onsager_even} and \eqref{eq:onsager_odd} to Eq. \eqref{eq:dc_conductivity_linear_q}, we obtain the time-parity sectors. 
The time-even sector is given by 
\begin{align}
\sigma^A_{ab,c}
=&-\frac{ie^2}{8\pi\hbar}
\int d\varepsilon\,
\partial_\varepsilon f(\varepsilon)
\operatorname{Tr}\Big[
\Delta\hat G\{\hat r_{c},\hat v_{a}\}\Delta\hat G\hat v_b
\nonumber\\
&-\Delta\hat G\{\hat r_{c},\hat v_{b}\}\Delta\hat G\hat v_a
\Big].
\label{eq:sigma_even}
\end{align}

Conversely, the time-odd part is 
\begin{align}
\sigma^S_{ab,c}
=&\frac{ie^2}{4\pi\hbar}
\operatorname{Re}\int d\varepsilon\,f(\varepsilon)
\operatorname{Tr}\Big[
\Delta\hat G\{\hat r_{c},\hat v_{a}\}
\partial_\varepsilon(\hat G^R+\hat G^A)\hat v_b
\nonumber\\
&+\Delta\hat G\{\hat r_{c},\hat v_{b}\}
\partial_\varepsilon(\hat G^R+\hat G^A)\hat v_a
\Big].
\label{eq:sigma_odd}
\end{align}

Equations \eqref{eq:sigma_even} and \eqref{eq:sigma_odd}, together with Eqs. \eqref{eq:magnetoelectric_tensor}, \eqref{odd-magnetoelectric} and \eqref{even-magnetoelectric}, describe the nonequilbrium magnetization response to a static field of any bulk noninteracting system. The time-even sector $\sigma^{\text{A}}_{ab,c}$ depends only on the spectral function at the Fermi level and is therefore a genuine Fermi-surface contribution. On the other hand, the time-odd sector $\sigma^{\text{S}}_{ab,c}$ has both Fermi-surface and Fermi-sea contributions (see Appendix \ref{Ap:Streda_Decomposition} and discussion from Ref. \cite{Ocana_23}).
In a clean crystal this sector describes the intrinsic orbital magnetoelectric response. In a disordered system, however, the time-odd sector should not in general be identified with a strictly disorder-independent intrinsic contribution: scattering can renormalize this sector in a nontrivial fashion through self-energy and vertex processes even when the time-even response is symmetry forbidden. Surprisingly, as shown in what follows and in the accompanying paper \cite{Pinho_PRL}, even weak disorder lead to remarkable modifications in the magneto-electric response, which defies the naive expectations.

\section{Results} 
\label{sec:VI-Results}

\begin{figure}[!t]
\centering
\includegraphics[width=\linewidth]{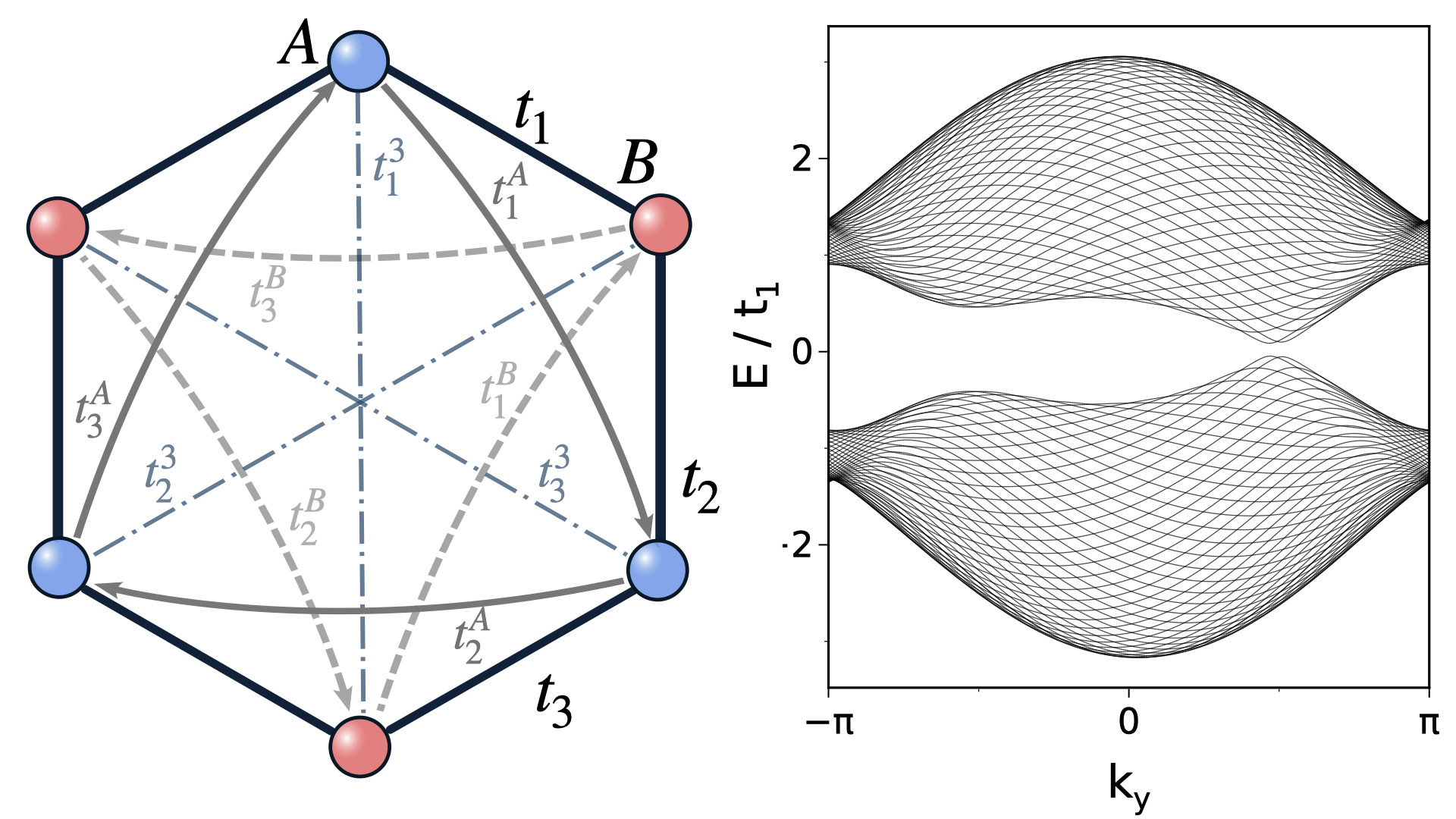}
\caption{Anisotropic Haldane model with $C_1$ symmetry. Left: Honeycomb plaquette depicting the  inequivalent nearest-neighbor  hoppings $t_j$, sublattice-dependent next-nearest-neighbor (NNN) hoppings $t_j^{A,B}$ and third-nearest neighbor hoppings $t_j^{3}$, with $j=1,2,3$ labeling the three bond orientations. For $\phi=\pi/2$, the complex NNN hopping amplitudes are $t_j^{A/B}=u_j^{A/B}+i\nu_{j}h_j^{B}$ where $\nu_{j}=\pm1$ sets the hopping chirality. We set, in units of $t_1$, $t_{\{1,2,3\}}=\{1,0.76,1.22\}$, $u^A_{\{1,2,3\}}=\{0.06,-0.025,0.02\}$,
$h^A_{\{1,2,3\}}=\{0.04,0.06,0.025i\}$, $u^B_{\{1,2,3\}}=\{-0.035,0.05,-0.015\}$,  
$h^B_{\{1,2,3\}}=\{0.055,0.03,0.05\}$,  $t^3_{\{1,2,3\}}=\{0.1,-0.035,0.065\}$ and $M=0.13$. Right: Clean band structure of the model. Each line is a slice along $k_y$ for a fixed $k_x$ value on the FBZ.}
\label{Fig1} 
\end{figure}

As a minimal testbed for the real-space formulation, we consider an anisotropic extension of the Haldane model on the honeycomb lattice. The Hamiltonian is

\begin{align}
H ={}&
-\sum_{\langle ij\rangle}
t_{ij}\,c_i^\dagger c_j
-\sum_{\langle\!\langle ij\rangle\!\rangle}
\left(
u_{ij}
+h_{ij}e^{i\xi_i\nu_{ij}\phi_H}
\right)c_i^\dagger c_j
\nonumber\\
&-
\sum_{\langle\!\langle\!\langle ij\rangle\!\rangle\!\rangle}
t^{(3)}_{ij}\,c_i^\dagger c_j
+\sum_i
\Delta_i\,c_i^\dagger c_i
+\mathrm{H.c.},
\label{eq:haldane_model}
\end{align}
where \(\Delta_i=\xi_iM+ w_i\), $\xi_i=\pm1$ labels the two sublattices, $w_i\in [-W/2,W/2]$ is a random scalar potential drawn from an uniform distribution, $W$ is the disorder strength and   
\(\nu_{ij}=\pm1\) sets the chirality of the complex second-neighbor
hopping. Here, the hopping amplitudes are allowed to depend on bond
orientation and sublattice, thereby lowering the point-group symmetry
of the standard Haldane model. The hoppings used in the calculations are depicted in Fig. \ref{Fig1}, and their numerical values stated in the caption.  For this choice of parameters, the model exhibits a small gap around   $\mu=0.0$ (see right panel of Fig. \ref{Fig1}). 

The anisotropy built-in in our toy model is motivated by symmetry (for a realistic model of a disordered crystal see the see accompanying paper \cite{Pinho_PRL}). In the conventional Haldane model, the threefold rotation \(C_{3z}\) forbids the coupling of an in-plane electric field to an out-of-plane magnetization. Since \(M_z\) is invariant under \(C_{3z}\), whereas \((E_x,E_y)\) transforms as an in-plane vector, symmetry requires \(K_{zx}=K_{zy}=0\). By making symmetry-related hopping amplitudes nonequivalent, the present model lowers the point group to \(C_1\), for which these components are symmetry allowed (See Table \ref{tab:ME_symmetry}).

\begin{table}[t]

\centering
\begin{ruledtabular}
\begin{tabular}{c c c}
Point group
&
\(K\)
&
\(M_z\) from \(E_{x,y}\)
\\
\hline
\(C_{3z}\)
&
\(
\begin{pmatrix}
0 & 0 & 0\\
0 & 0 & 0\\
0 & 0 & 0
\end{pmatrix}
\)
&
Forbidden
\\[4mm]
\(C_1\)
&
\(
\begin{pmatrix}
0 & 0 & 0\\
0 & 0 & 0\\
\mathbf{K_{zx}} & \mathbf{K_{zy}} & 0

\end{pmatrix}
\)
&
Allowed

\end{tabular}
\end{ruledtabular}

\caption{
Symmetry-allowed orbital magnetoelectric response \(M_i=K_{ij}E_j\) for the strictly two-dimensional geometry considered here. Since the orbital magnetization generated by in-plane currents is restricted to the out-of-plane component \(M_z\), only the coefficients \(K_{zx}\) and \(K_{zy}\) are relevant. Threefold rotational symmetry \(C_{3z}\) forbids both coefficients, so the linear magnetoelectric response vanishes. When the symmetry is lowered to \(C_1\), both couplings between an in-plane electric field and an out-of-plane orbital magnetization are allowed.
}
\label{tab:ME_symmetry}
\end{table}

\begin{figure}[!t]

\centering
\includegraphics[width=\linewidth]{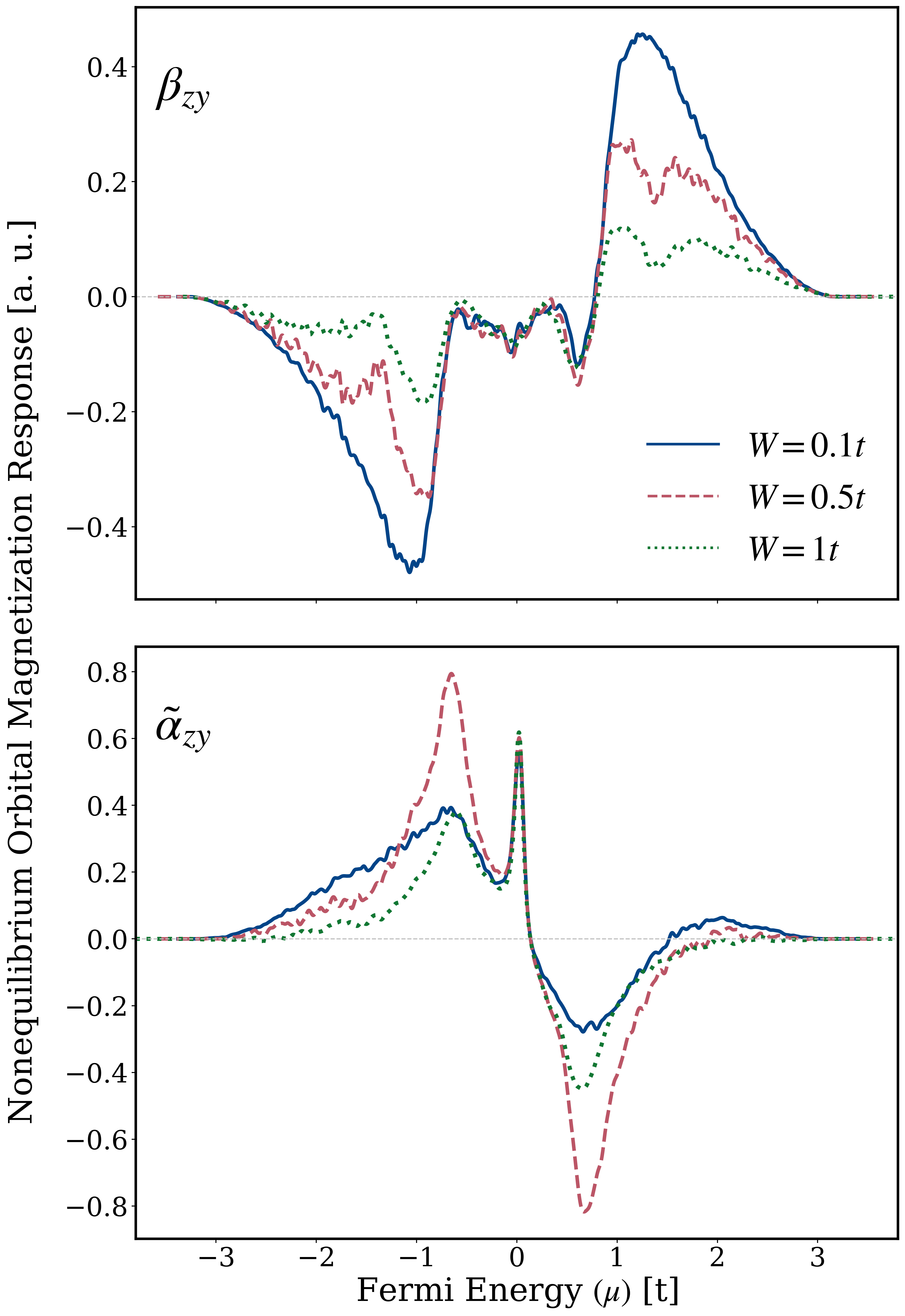}
\caption{Comparison of the nonequilibrium orbital magnetization response of the distorted Haldane model in the presence of onsite disorder. The top and bottom panels show the time-even response $\beta_{zy}$ and time-odd response $\tilde{\alpha}_{zy}$, respectively. The blue dashed line, red solid and green dotted lines correspond to disorder strengths $W=0.1t$, $W=0.5t$ and $W=1t$, respectively.}
\label{Fig2} 
\end{figure}

We now present our numerical results. Figure \ref{Fig2} compares the time-even and time-odd sectors of the magnetoelectric response obtained for different disorder strengths. The two Onsager sectors are seen to display markedly different evolutions with increasing disorder. The time-even contribution, $\beta_{zy}$, is progressively suppressed as the disorder strength is increased, with its main spectral features becoming substantially reduced for the largest disorder strength considered ($W = t$). This is the typical behavior of a Fermi-surface transport quantity, which is expected to display monotonically decreasing behavior with the disorder strength. Strikingly, the time-odd contribution $\tilde{\alpha}_{zy}$ exhibits a non-monotonic dependence on disorder. In particular, the outer spectral features are initially enhanced at intermediate disorder strength before being suppressed again for stronger disorder. This behavior demonstrates that the time-odd response is not governed by a simple disorder-broadening mechanism (as often assumed), but instead reflects a nontrivial reconstruction of the underlying quantum response. The suppression observed at the largest disorder strength is consistent with the onset of strong-disorder effects, which we attribute to incipient localization and the associated reduction of extended-state transport. 

 An interesting direction for future work is the investigation of current-induced orbital magnetization in the distorted Haldane model using the momentum-space perturbative methods put forward in Refs. \cite{rouKineticOrbitalMoments2017,sahin_pancharatnam-berry_2018-1,Liu_24,Veneri_25,sarkar_extrinsic_2026,cullen_theory_2026}. In the weak-disorder regime, $W \ll t$, we expect the results obtained with our exact real-space treatment to quantitatively agree with the prediction of a self-consistent microscopic theory that consistently accounts for the position operator in the Bloch representation. For the symmetric Anderson disorder considered in this work, conventional skew-scattering processes are described by fourth-order scattering diagrams and thus are sub-leading in a weak disorder (Born) expansion \cite{Ado_2015,Milletari_16,Konig_16,Milletari_17,
Perkins2024_review}. This is in contrast with more realistic models of point defects and short-range impurities, where non-vanishing third-order impurity diagrams provide the leading contribution to orbital skew scattering  \cite{Veneri_25}. As such, one may expect other extrinsic mechanisms that manifest already at the first Born approximation level, such as the orbital side jump, to dominate the extrinsic OM response of the model studied here in the small $W$ limit.

\section{Conclusions} 
\label{sec:VII-Conclusions}

A comprehensive description of nonequilibrium OM in real materials requires an approach capable of treating quantum geometry and disorder on equal footing. In this work, we have introduced a real-space framework for simulating dc electric-field induced OM that treats spatial inhomogeneities in an exact fashion to achieve this objective. Our formulation is grounded in a basis-independent and gauge-invariant description of non-equilibrium OM [Eqs. (\ref{eq:dc_conductivity_linear_q}), (\ref{eq:sigma_even}) and \ref{eq:sigma_odd})], which we have introduced here.
The present formulation extends the theory of electric field- and current-induced orbital magnetization beyond the crystalline limit. Its central object is the conductivity tensor linear in wavevector rather than a direct expectation value of an orbital-moment operator. This distinction is important in a conducting nonequilibrium state, where  $\langle \mathbf r\times\mathbf v \rangle $ is a ill-defined concept which depends on the arbitrary choice of real-space origin. By extracting the magnetic-dipole contribution from the spatially dispersive current response, the present construction provides a sound definition of non-equilibrium orbital magnetization that remains meaningful in bulk inhomogeneous systems.  In a pristine crystal, the real-space Kubo kernel reduces exactly to the conventional momentum-space response. Its advantage appears once translational invariance is lost: disorder and spatial inhomogeneity enter directly through the Hamiltonian, with no need to impose a supercell or to rely on perturbative quantum-kinetic or diagrammatic treatments of impurity scattering. The application of the formalism to the anisotropic Haldane model shows that disorder dramatically changes the non-equilibrium orbital magnetization even in the weak-disorder limit. This is an important finding which indicates that extrinsic mechanisms may overcome the intrinsic magneto-electric response in realistic materials. This intuition is validated for a cuprate model system with point defects in the accompanying paper \cite{Pinho_PRL}. 

Our real-space theory thus complements the existing Bloch-space formulations by highlighting the crucial role played by extrinsic charge-to-orbital conversion mechanisms stemming from disorder. The Onsager decomposition of the magneto-electric tensor also sheds light on the relation between symmetry and microscopic origin. The time-even sector of the magneto-electric response is a genuine Fermi-surface effect, hence governed by the interplay of disorder and quantum geometry. The time-odd sector contains the clean (intrinsic) orbital magnetoelectric response but, once disorder is present, it generally acquires significant scattering-induced corrections. We have therefore reserved the term ``intrinsic'' for the clean crystalline contribution and have used time-even/odd exclusively as symmetry labels in the disordered problem.
Although illustrated here on a toy model with Anderson disorder, our framework is equally applicable to realistic disorder models \cite{Pinho_PRL}. 

In this first investigation, the focus was on the orbital magneto-electric response of a bulk medium with short-range disorder. In future work, it would be interesting to extend our real-space framework to interfaces, quasiperiodic structures, and spatially varying magnetic or orbital textures. Our approach opens a route to nonequilibrium orbital magnetism in settings that are inaccessible to conventional Bloch-space methods.

\section{Acknowledgements}
J.M.A.P., B.A., H.P.V. and J.M.V.P.L. acknowledge support from Fundação para a Ciência e a Tecnologia (FCT, Portugal) through
the Strategic Funding UID/04650/2025, Grant No. UID/PRR/04650/2025 and Grant No. UID/PRR2/04650/2025. Further support from Fundação para a Ciência e a Tecnologia
(FCT, Portugal) is acknowledged through the project QPI-MoireS No. 2023.17745.ICDT (B.A.,
J.M.A.P., J. M. V. P. L. and H.P.V) and Grants
No. 2023.02155.BD (J.M.A.P.) and No. 2024.00560.BD (H.V.P). H.P.V. acknowledges computational resources from the Laboratory for
Advanced Computing at the University of Coimbra. A.F.,
J.M.V.P.L., J.M.A.P. and H.P.V. acknowledge computational
resources from the University of York High-Performance Computing service,
Viking. D.P. acknowledges funding from Horizon Europe under the
Marie Skłodowska-Curie grant agreement No.101206626. I.S. acknowledges
support from Grant No.~PID2025-171272NB-I00 funded by
MCIN/AEI/10.13039/501100011033. A.V. thanks
Roberto Raimondi for fruitful discussions.

\bibliographystyle{naturemag}
\bibliography{ref}

\appendix

\onecolumngrid

\section{Linearized  current vertex Derivation} \label{Ap:current_vertex}
Here we present the derivation of the current vertex for a dispersive field and prove that, for both continuum and tight-binding Hamiltonians, we obtain Eq. \eqref{eq:vq_expansion} from the main text. 
\subsection{Continuum Hamiltonian}
Let us first consider a continuum Hamiltonian. Minimal coupling to the external electromagnetic field yields 

\begin{equation}
H\left(t\right)=-\frac{\hbar^{2}}{2m}\left(\mathbf{\hat{p}}+\mathbf{A}\left(\mathbf{\hat{r}},t\right)\right)^{2}+V(\mathbf{\hat{r}}),
\end{equation}

where $\mathbf{\hat{r}}$ and $\mathbf{\hat{p}}$ are the position and momentum operators, respectively. Expanding in $\mathbf{A}$ yields
\begin{equation}
H\left(t\right)=-\frac{\hbar^{2}}{2m}\mathbf{\hat{p}}^{2}-\frac{\hbar^{2}}{2m}\left(\mathbf{\hat{p}} \cdot\mathbf{A}\left(\mathbf{\hat{r}},t\right)+\mathbf{A}\left(\mathbf{\hat{r}},t\right) \cdot \mathbf{\hat{p}}\right)-\frac{\hbar^{2}}{2m}\mathbf{A}^{2}\left(\mathbf{\hat{r}},t\right)+V(\mathbf{\hat{r}}).
\label{eq:continuum_exp}
\end{equation}

Next, we expand the gauge potential in Fourier space as 

\begin{equation}
A^{a}\left(\mathbf{\hat{r}},t\right)=\int\mathbf{q}A^{a}\left(\mathbf{q},t\right)e^{i\mathbf{q}\cdot\mathbf{\hat{r}}},
\label{eq:Fourier_Exp}
\end{equation}

where $\mathbf{A}(\mathbf{q},t)$ is now a number. Inserting Eq. \eqref{eq:Fourier_Exp} into Eq. \eqref{eq:continuum_exp} and using the definition of paramagnetic operator in Eq. \eqref{eq:perturbation_expansion} allows us to obtain the paramagnetic operator as 
\begin{equation}
\hat{v}_{a}\left(\mathbf{q}\right)=-\frac{\hbar^{2}}{2m}\left\{ \mathbf{\hat{p}},e^{i\mathbf{q}\cdot\mathbf{\hat{r}}}\right\}.
\label{finiteq_continuum}
\end{equation}
Likewise, the diamagnetic operator is given by
\begin{equation}
\hat{\Delta}_{ab}\left(\mathbf{q},\mathbf{q}'\right)=-\frac{\hbar^{2}}{2m}\delta_{ab}e^{i\left(\mathbf{q}+\mathbf{q}'\right)\cdot\mathbf{\hat{r}}}.
\label{eq:Diamagnetic_continuum}
\end{equation}

Expanding Eq. \eqref{finiteq_continuum} to first order in $\mathbf{q}$ and using that the velocity operator of the unpertrubed system is $\hat{v}_i=\hat{p}_i/m$, we regain Eq. \eqref{eq:vq_expansion}.

\subsection{Tigh-Binding Hamiltonian}
Now onto tight-binding Hamiltonians, which in all generality can be expressed
in real-space as 
\begin{equation}
H_{0}=\sum_{\alpha\beta}h_{\alpha\beta}\hat c^{\dagger}_{\alpha}\hat c_{\beta},
\end{equation}
where $\alpha$ and $\beta$ label atomic orbitals located at positions $\mathbf{R_{\alpha}}$
and $\mathbf{R_{\beta}}$, The hopping terms correspond to $h_{\alpha\beta}$
and the coupling to the external electromagnetic field is introduced
through the usual Peierls substitution

\begin{equation}
H_{A}(t)=\sum_{\alpha,\beta}e^{-i\frac{e}{\hbar}\int^{\mathbf{R}_{\beta}}_{\mathbf{R}_{\alpha}}\mathbf{A}(\mathbf{r},t)\cdot d\mathbf{s}}h_{\alpha\beta}\hat c^{\dagger}_{\alpha}\hat c_{\beta},\label{Definition_H_A}
\end{equation}
where $e$ denotes the elementary electron charge, $d\mathbf{s}$ is the infinitesimal curve connecting the two atomic positions $\mathbf{R_{\alpha}}$ and $ \mathbf{R_{\beta}}$ and the vector potential,
$\mathbf{A}(\mathbf{r},t)$, is described in the Weyl gauge as
\begin{equation}
\partial_{t}\mathbf{A}(\mathbf{r},t)=-\mathbf{E}(\mathbf{r},t),\label{eq:velocity_gauge}
\end{equation}
where $\mathbf{E}$ is the applied electric field.

Assuming small perturbations, we expand Eq.$\,$\eqref{Definition_H_A}
up to second order in the vector potential,

\begin{equation}
H^{(2)}_{A}(t)=\sum_{\alpha,\beta}\left[1-i\frac{e}{\hbar}\int^{\mathbf{R}_{\beta}}_{\mathbf{R}_{\alpha}}\mathbf{A}(\mathbf{r},t)\cdot d\mathbf{s}-\frac{e^{2}}{2\hbar ^ 2 }\left(\int^{\mathbf{R}_{\beta}}_{\mathbf{R}_{\alpha}}\mathbf{A}(\mathbf{r},t)\cdot d\mathbf{s}\right)^{2}+\cdots\right]h_{\alpha\beta}\hat c^{\dagger}_{\alpha}\hat c_{\beta}.\label{eq:pertubration}
\end{equation}
The linear term yields the paramagnetic current operator, while
the quadratic contribution generates the diamagnetic response required
to preserve gauge invariance. Following this result, we express the
vector potential in terms of its spatial Fourier transform,

\begin{equation}
\mathbf{A}(\mathbf{r},t)=\int d\mathbf{q}\,\mathbf{A}(\mathbf{q},t)e^{i\mathbf{q}\cdot\mathbf{r}},
\end{equation}
which allows one to write the Peierls phase as

\begin{align}
\int^{\mathbf{R}_{\beta}}_{\mathbf{R}_{\alpha}}\mathbf{A}(\mathbf{r},t)\cdot d\mathbf{s} & =\int^{\mathbf{R}_{\beta}}_{\mathbf{R}_{\alpha}}\int d\mathbf{q}e^{i\mathbf{q}\cdot\mathbf{r}}\mathbf{A}(\mathbf{q},t)\cdot d\mathbf{s}=\int d\mathbf{q}\frac{e^{i\mathbf{q}\cdot\mathbf{R}_{\beta}}-e^{i\mathbf{q}\cdot\mathbf{R}_{\alpha}}}{i\mathbf{q}\cdot\Delta\mathbf{R}_{\alpha\beta}}\mathbf{A}(\mathbf{q},t)\cdot\Delta\mathbf{R}_{\alpha\beta},
\label{eq:peierlsphase}
\end{align}
where $\Delta\mathbf{R}_{\alpha\beta}=\mathbf{R}_{\beta}-\mathbf{R}_{\alpha}.$ Our interest covers electromagnetic fields slowly varying  on the
scale of atomic distances, i.e., with long wavelengths, $q|\Delta\mathbf{R}_{\alpha\beta}|\ll1$, wherein the vector potential may
be regarded as approximately constant along the integration contour.
Therefore, we can approximate

\begin{equation}
\int^{\mathbf{R}_{\beta}}_{\mathbf{R}_{\alpha}}\mathbf{A}(\mathbf{r},t)\cdot d\mathbf{s}\simeq\mathbf{A}\left(\frac{\mathbf{R}_{\alpha}+\mathbf{R}_{\beta}}{2},t\right)\cdot\Delta\mathbf{R}_{\alpha\beta},
\end{equation}
which can be expressed in Fourier space as
\begin{equation}
\int^{\mathbf{R}_{\beta}}_{\mathbf{R}_{\alpha}}\mathbf{A}(\mathbf{r},t)\cdot d\mathbf{s}\simeq\int d\mathbf{q}e^{\frac{i}{2}\mathbf{q}\cdot(\mathbf{R}_{\alpha}+\mathbf{R}_{\beta})}\mathbf{A}(\mathbf{q},t)\cdot\Delta\mathbf{R}_{\alpha\beta}.
\end{equation}

The total Hamiltonian takes the form

\begin{equation}
\hat H_{A}=\hat H_{0}+e\int d\mathbf{q}\,\hat v_{a}(\mathbf{q})A_{a}(\mathbf{q})+e ^ 2\int d\mathbf{q}d\mathbf{q}'\hat \Delta_{ab}(\mathbf{q},\mathbf{q}')A_{a}(\mathbf{q})A_{b}(\mathbf{q}'),
\end{equation}
explicitly defining the operators coupled to the powers of the vector potential, i.e., the paramagnetic current operator
\begin{equation}
\hat v_{a}(\mathbf{q})=-\frac{i}{\hbar}\sum_{\alpha,\beta}e^{\frac{i}{2}\mathbf{q}\cdot(\mathbf{R}_{\alpha}+\mathbf{R}_{\beta})}\Delta R^{a}_{\alpha\beta}h_{\alpha\beta}\hat c^{\dagger}_{\alpha}\hat c_{\beta},\label{eq:paramagnetic}
\end{equation}
and the diamagnetic current operator
\begin{equation}
\hat \Delta_{ab}(\mathbf{q},\mathbf{q}')=-\frac{1}{2\hbar ^ 2}\sum_{\alpha,\beta}e^{\frac{i}{2}(\mathbf{q}+\mathbf{q}')\cdot(\mathbf{R}_{\alpha}+\mathbf{R}_{\beta})}\Delta R^{a}_{\alpha\beta}\Delta R^{b}_{\alpha\beta}h_{\alpha\beta}\hat c^{\dagger}_{\alpha}\hat c_{\beta}.\label{eq:diamagnetic}
\end{equation}
These objects, in the uniform-field limit, reduce to
\begin{equation}
\begin{alignedat}{1}\lim_{\mathbf{q}\rightarrow 0}\hat v_{a}\left(\mathbf{q}\right) & =\hat v_{a},\\
\lim_{\mathbf{q},\mathbf{q}'\rightarrow0}\hat \Delta_{ab}\left(\mathbf{q},\mathbf{q}'\right) & =\hat \Delta_{ab},
\end{alignedat}
\end{equation}
with 
\begin{equation}
\begin{aligned}\hat v_{a} & =-\frac{i}{\hbar}\sum_{\alpha,\beta}\Delta R^{a}_{\alpha\beta}h_{\alpha\beta}\hat c^{\dagger}_{\alpha}\hat c_{\beta}=-\frac{i}{\hbar}[H_0,\hat{r}^a],\\
\hat \Delta_{ab}= & -\frac{1}{2 \hbar^2}\sum_{\alpha,\beta}\Delta R^{a}_{\alpha\beta}\Delta R^{b}_{\alpha\beta}h_{\alpha\beta}\hat c^{\dagger}_{\alpha}\hat c_{\beta}=-\frac{1}{2\hbar^2}[[H_0,\hat r^a],\hat r^b],
\end{aligned}
\label{eq:uniform}
\end{equation}
and $\hat{\mathbf{r}}=\sum_i \mathbf{R}_i \hat{c}_i^{\dagger}\hat{c}_i$ is the position operator.

It is useful to recast the operators in Eqs. (\ref{eq:paramagnetic}) and (\ref{eq:diamagnetic}) in terms of their uniform counterparts, Eqs. (\ref{eq:uniform}), as 

\begin{subequations}
\label{eq:vertices}
\begin{align}\hat v_{a}(\mathbf{q}) & =e^{\frac{i}{2}\mathbf{q}\cdot\mathbf{\hat{r}}}\hat v_{a}e^{\frac{i}{2}\mathbf{q}\cdot\mathbf{\hat{r}}},\label{eq:ops_position_op1}\\
\hat \Delta_{ab}(\mathbf{q},\mathbf{q}') & =e^{\frac{i}{2}\left(\mathbf{q}+\mathbf{q}'\right)\cdot\mathbf{\hat{r}}}\hat \Delta_{ab}e^{\frac{i}{2}\left(\mathbf{q}+\mathbf{q}'\right)\cdot\mathbf{\hat{{r}}}}.\label{eq:ops_position_op2}
\end{align}
\end{subequations}

It can be easily seen that expanding Eq. \eqref{eq:ops_position_op1} yields Eq. \eqref{eq:vq_expansion}
 from the main text. 

\section{Linear Response Theory} \label{Ap:Linear_Response_Theory}

The linear-response current under the driving of an inhomogeneous vector potential can be expressed as
\begin{equation}
J_{a}(\omega,\mathbf{q})=\int d\mathbf{q}'\Phi_{ab}(\omega,\mathbf{q},-\mathbf{q}')A_{b}(\omega,\mathbf{q}'),
\end{equation}
where the response kernel is given by the
Kubo formula

\begin{equation}
\Phi_{ab}(\omega,\mathbf{q},\mathbf{q}')=\langle\Delta_{ab}(\mathbf{q},\mathbf{q}')\rangle_{0}+\Pi_{ab}(\omega,\mathbf{q},\mathbf{q}'),\label{eq:kernel_generic}
\end{equation}
with

\begin{equation}
\Pi_{ab}(\omega,\mathbf{q},\mathbf{q}')=-\frac{ie^2}{\hbar}\int^{\infty}_{0}dte^{i(\omega+i0^{+})t}\left\langle [\hat v_{a}(\mathbf{q},t),\hat v_{b}(\mathbf{q}',0)]\right\rangle _{0}.
\label{eq:general_2presp}
\end{equation}
The ensemble average is represented by the symbol $\langle... \rangle_0\equiv\mathrm{Tr}\{...\rho_0 \}$, where $\rho_{0}=\exp(-\beta H_0)/\mathcal{Z}$ is the unperturbed density matrix, and the trace runs over many-particle states. The formal result \ref{eq:general_2presp} is general. However, by virtue of the single-particle nature of the present problem, the expression above can be re-organized in terms of the  unperturbed \textit{single particle Hamiltonian} eigenbasis, obtaining
\begin{align*}
\Pi_{ab}(\omega,\mathbf{q},\mathbf{q}')=
\frac{e^2}{\hbar}\sum_{\lambda\lambda'}\Bigg[ & \frac{f_{\lambda}}{\varepsilon_{\lambda}-\varepsilon_{\lambda'}+\omega+i0^{+}}v^{\lambda\lambda'}_{a}(\mathbf{q})v^{\lambda'\lambda}_{b}(\mathbf{q}')+\frac{f_{\lambda}}{\varepsilon_{\lambda}-\varepsilon_{\lambda'}-\omega-i0^{+}}v^{\lambda\lambda'}_{b}(\mathbf{q}')v^{\lambda'\lambda}_{a}(\mathbf{q})\Bigg]
\end{align*}
where $|\lambda\rangle$ and $|\lambda'\rangle$ are the eigenstates of the system and $f_\lambda$ the Fermi-Dirac distribution.
From the introduction of a Dirac delta function over the first energy
and by defining the retarded ($G^{R}$) and advanced ($G^{A}$) Green's
functions as 
\begin{equation}
G^{R/A}_{\lambda\lambda'}(\varepsilon)
=
\frac{
\delta_{\lambda\lambda'}
}{
\varepsilon-\varepsilon_{\lambda}\pm i 0^+
},
\end{equation}
the response function may be reorganized as 
\begin{align*}
\Pi_{ab}(\omega,\mathbf{q},\mathbf{q}')=
\frac{e^2}{\hbar}\sum_{\lambda\lambda'}\int d\varepsilon f(\varepsilon)\Big[ & \delta(\varepsilon-\varepsilon_{\lambda})v^{\lambda\lambda'}_{a}(\mathbf{q})G^{R}_{\lambda'\lambda'}(\varepsilon+\omega)v^{\lambda'\lambda}_{b}(\mathbf{q}')+\delta(\varepsilon-\varepsilon_{\lambda})v^{\lambda\lambda'}_{b}(\mathbf{q}')G^{A}_{\lambda'\lambda'}(\varepsilon-\omega)v^{\lambda'\lambda}_{a}(\mathbf{q})\Big].
\end{align*}
or in an equivalent basis-independent formulation as
\begin{equation}
\Pi_{ab}(\omega,\mathbf{q},\mathbf{q}')=
\frac{e^2}{\hbar}\int d\varepsilon f(\varepsilon)\mathrm{Tr}\Big[\delta(\varepsilon-\hat H)\hat v_{a}(\mathbf{q})\hat G^{R}(\varepsilon+\omega)\hat v_{b}(\mathbf{q}')+\delta(\varepsilon-\hat H)\hat v_{b}(\mathbf{q}')\hat G^{A}(\varepsilon-\omega)\hat v_{a}(\mathbf{q})\Big],
\label{eq:Pi_general}
\end{equation}
which can be computed efficiently with real-space numerical methods based on Chebyshev expansions \citep{weise_kernel_2006,Ferreira_15,joao_kite_2020}. Here and henceforth, trace formulas should be understood as traces per physical volume  $\Omega$. The complete linear-response kernel is obtained by inserting Eq. \eqref{eq:Pi_general} into Eq. \eqref{eq:kernel_generic}. Taking $\mathbf{q'}=\mathbf{-q}$ and using that the diamagnetic operator $\Delta(\mathbf{q},\mathbf{-q})$ is momentum independent as per Eqs. \eqref{eq:Diamagnetic_continuum} and \eqref{eq:ops_position_op2} gives us Eq. \eqref{eq:kernel_generic} from the main text.

\section{dc limit Derivation} \label{Ap:dc_limit_derivation}
Differentiating Eq. \eqref{eq:Phi_ijc_trace} with respect to the frequency,
and subsequently setting $\omega=0$, gives 
\begin{align}
\partial_{\omega}\Phi_{ab,c}(0)
&=
\frac{ie^2}{2\hbar}
\int d\varepsilon\,
f(\varepsilon)\,
\mathrm{Tr}\Big[
\delta(\varepsilon-\hat H)
\{\hat r_{c},\hat v_{a}\}
\nonumber
\partial_{\varepsilon}\hat G^{R}(\varepsilon)
\hat v_{b}
\nonumber\\
&\quad\qquad\qquad\qquad\qquad-
\delta(\varepsilon-\hat H)
\hat v_{a}
\partial_{\varepsilon}\hat G^{R}(\varepsilon)
\{\hat r_{c},\hat v_{b}\}
\nonumber\\
&\quad\qquad\qquad\qquad\qquad+
\delta(\varepsilon-\hat H)
\{\hat r_{c},\hat v_{b}\}
\partial_{\varepsilon}\hat G^{A}(\varepsilon)
\hat v_{a}
\nonumber\\
&\quad\qquad\qquad\qquad\qquad-
\delta(\varepsilon-\hat H)
\hat v_{b}
\partial_{\varepsilon}\hat G^{A}(\varepsilon)
\{\hat r_{c},\hat v_{a}\}
\Big],
\label{eq:derivative_phi}
\end{align}
where we have used 
\begin{equation}
\partial_{\omega}\hat G^{R}(\varepsilon+\omega)=\partial_{\varepsilon}\hat G^{R}(\varepsilon+\omega),\qquad\partial_{\omega}\hat G^{A}(\varepsilon-\omega)=-\partial_{\varepsilon}\hat G^{A}(\varepsilon-\omega).
\end{equation}
Using Hermitian conjugation in tandem with the cyclic property of the
trace, one finds that 
\begin{equation}
\mathrm{Tr}\left[\delta(\varepsilon-\hat H)\hat v_{b}\partial_{\varepsilon}\hat G^{A}(\varepsilon)\{\hat r_{c},\hat v_{a}\}\right]=\mathrm{Tr}\left[\delta(\varepsilon-\hat H)\{\hat r_{c},\hat v_{a}\}\partial_{\varepsilon}\hat G^{R}(\varepsilon)\hat v_{b}\right]^{*}.\label{eq:conjugation_identity}
\end{equation}
This results recasts Eq. \eqref{eq:derivative_phi} as 
\begin{align}
\partial_{\omega}\Phi_{ab,c}(0)=-\frac{e^2}{\hbar}\int d\varepsilon\,f(\varepsilon)\,\mathrm{Im}\,\mathrm{Tr}\Big[ & \delta(\varepsilon-\hat H)\{\hat r_{c},\hat v_{a}\}\partial_{\varepsilon}\hat G^{R}(\varepsilon)\hat v_{b}-\delta(\varepsilon-\hat H)\hat v_{a}\partial_{\varepsilon}\hat G^{R}(\varepsilon)\{\hat r_{c},\hat v_{b}\}\Big].\label{eq:phi_derivative_compact}
\end{align}

Using $\sigma_{ij}=\Phi_{ij}/i\omega$ and \eqref{eq:phi_derivative_compact},
the dc conductivity tensor associated with the linear-in-$q$ response becomes 
\begin{equation}
\sigma_{ab,c}=\frac{ie^2}{\hbar}\int d\varepsilon\,f(\varepsilon)\,\mathrm{Im}\,\mathrm{Tr}\Big[\delta(\varepsilon-\hat H)\{\hat r_{c},\hat v_{a}\}\partial_{\varepsilon}\hat G^{R}(\varepsilon)\hat v_{b}-\delta(\varepsilon-\hat H)\hat v_{a}\partial_{\varepsilon}\hat G^{R}(\varepsilon)\{\hat r_{c},\hat v_{b}\}\Big],
\end{equation}

which can be written purely in terms of Green's functions as
\begin{equation}
\sigma_{ab,c}=\frac{ie^2}{2\pi\hbar}\int d\varepsilon\,f(\varepsilon)\,\mathrm{Re}\,\mathrm{Tr}\Big[\Delta \hat G\left(\varepsilon\right)\{\hat r_{c},\hat v_{a}\}\partial_{\varepsilon}\hat G^{R}(\varepsilon)\hat v_{b}-\Delta \hat G\left(\varepsilon\right) \hat v_{a}\partial_{\varepsilon}\hat G^{R}(\varepsilon)\{\hat r_{c},\hat v_{b}\}\Big].
\label{eq:supp_dc_conductivity_linear_q}
\end{equation}

\section{Crystalline Bloch limit} \label{Ap:Crystalline_Bloch_Limit}

Having derived a real-space expression for the linear-response kernel,
it is important to demonstrate that it reproduces the established
momentum-space formalism in the limit of translational invariance.
In this section, we show that our expression reduces exactly to the
response function derived for crystalline systems in$\,$\citep{shinadaOrbitalOpticalActivity2024},
thereby establishing the equivalence between the two formulations.

We will focus on the paramagnetic contribution as the diamagnetic one trivially follows with an analogous treatment. For periodic systems, the linear-response paramagnetic contribution may be written as
\begin{equation}
\begin{aligned}\Pi^{\text{Bloch}}_{ab}(\boldsymbol{q},\omega) & =\int d\varepsilon d\mathbf{k}f(\varepsilon)\text{Tr}_{\mathbf{k}}\left[\Delta \hat G(\boldsymbol{k}-\frac{\mathbf{q}}{2}, \omega)\hat v^{a}_{k} \hat G^{R}(\boldsymbol{k}+\frac{\mathbf{q}}{2},\varepsilon+\omega) \hat v^{b}_{k}\right.\\
 & \left.\qquad\qquad\qquad\,\,\,\,\,\,\,\,\,\,\,+\hat G^{A}(\boldsymbol{k}-\frac{\mathbf{q}}{2},\varepsilon-\omega)\hat v^{a}_{k}\Delta \hat G(\boldsymbol{k}+\frac{\mathbf{q}}{2},\varepsilon)\hat v^{b}_{k}\right],
\end{aligned}
\label{eq:shinada_Eq}
\end{equation}
where $\hat v^{a}_{k}=\frac{\partial \hat H\left(\mathbf{k}\right)}{\partial k_{a}}$,
$\hat G^{R/A}(\boldsymbol{k},\varepsilon)=\left(\varepsilon-\hat H\left(\mathbf{k}\right)\pm i0^+\right)^{-1}$,
$\Delta \hat G(\boldsymbol{k},\varepsilon)=\hat G^{R}(\boldsymbol{k},\varepsilon)-\hat G^{A}(\boldsymbol{k},\varepsilon)$,
$\hat H\left(\mathbf{k}\right)$ is the Bloch Hamiltonian, and $\text{Tr}_{\mathbf{k}}$ denotes the trace over the internal degrees of freedom of the unit
cell. Obtained from Eq.~\eqref{eq:Pi_general} with $\mathbf{q}'=-\mathbf{q}$, our real-space expression for the paramagnetic contribution to the response kernel is
\begin{equation}
\Pi_{ab}(\omega,\mathbf{q})=\frac{1}{\hbar}\int d\varepsilon f(\varepsilon)\mathrm{Tr}\Big[\delta(\varepsilon-\hat H)\hat v_{a}(\mathbf{q})\hat G^{R}(\varepsilon+\omega)\hat v_{b}(-\mathbf{q})\ +\delta(\varepsilon-\hat H)\hat v_{b}(-\mathbf{q})\hat G^{A}(\varepsilon-\omega)\hat v_{a}(\mathbf{q})\Big]
\label{eq:Reference_Parm_Sup}
\end{equation}
To establish the connection with the Bloch formalism, we assume discrete
translational symmetry and evaluate the current operator in a Bloch
basis. For a periodic lattice, the Bloch basis is defined as 
\begin{equation}
\ket{\psi_{n}\left(\mathbf{k}\right)}=\frac{1}{\sqrt{\Omega}}\sum_{\mathbf{R}}e^{-i\mathbf{k}\cdot\left(\mathbf{R}+\mathbf{\delta}_{n}\right)}\ket{\mathbf{R}+\mathbf{\delta}_{n}}
\end{equation}
where $\mathbf{R}$ labels the Bravais lattice vectors, while $\mathbf{\delta}_{n}$
denotes the position of sublattice $n$ within the unit cell. The
matrix elements of the paramagnetic operator in Eq.\eqref{eq:paramagnetic} 
are therefore 
\begin{equation}
\bra{\psi_{n}\left(\mathbf{k}\right)}\hat v_{a}\left(\mathbf{q}\right)\ket{\psi_{m}\left(\mathbf{k}'\right)}=-\frac{ie}{\Omega}\sum_{\mathbf{R},\mathbf{R}'}e^{i\frac{\mathbf{q}}{2}\cdot\left(\mathbf{R}'+\mathbf{\delta}_{m}+\mathbf{R}+\mathbf{\delta}_{n}\right)}e^{i\mathbf{k}\cdot\left(\mathbf{R}+\mathbf{\delta}_{n}\right)}e^{-i\mathbf{k}'\cdot\left(\mathbf{R}'+\mathbf{\delta}_{m}\right)}\Delta R^{a}h^{nm}_{\Delta\mathbf{R}}
\end{equation}
with $\Delta\mathbf{R}=\mathbf{R}-\mathbf{R}'+\mathbf{\delta}_{n}-\mathbf{\delta}_{m}$. The Hamiltonian solely depends on the difference $\Delta\mathbf{R}$
and the sub lattices $n,m$ due to translation symmetry. Performing
the change of variable $\mathbf{R}'+\delta_{m}=\mathbf{R}+\delta_{n}-\Delta\mathbf{R}$
we can write 
\begin{equation}
\begin{aligned}\bra{\psi_{n}\left(\mathbf{k}\right)}\hat v_{a}\left(\mathbf{q}\right)\ket{\psi_{m}\left(\mathbf{k}'\right)} & =-\frac{ie}{\Omega}\sum_{\mathbf{R},\Delta\mathbf{R}}e^{i\mathbf{q}\cdot\left(\mathbf{R}+\delta_{n}\right)-i\frac{\mathbf{q}}{2}\cdot\Delta\mathbf{R}}e^{i\mathbf{k}\cdot\left(\mathbf{R}+\mathbf{\delta}_{n}\right)}e^{-i\mathbf{k}'\cdot\left(\mathbf{R}+\delta_{n}-\Delta\mathbf{R}\right)}\Delta R^{a}h^{nm}_{\Delta\mathbf{R}}\\
 & =-ie\sum_{\Delta\mathbf{R}}\delta\left(\mathbf{k}+\mathbf{q}-\mathbf{k'}\right)e^{i\left(\mathbf{k}'-\frac{\mathbf{q}}{2}\right)\cdot\Delta\mathbf{R}}\Delta R^{a}h^{nm}_{\Delta\mathbf{R}}.
\end{aligned}
\end{equation}
The summation over lattice vectors immediately produces momentum conservation,
$\delta\left(\mathbf{k}+\mathbf{q}-\mathbf{k'}\right)$, showing that
the paramagnetic operator $v_{a}\left(\mathbf{q}\right)$ only
couples Bloch states whose crystal momenta differ by the transferred
momentum $\mathbf{q}$. For later convenience, we consider matrix
elements between the states $\ket{\psi_{n}\left(\mathbf{k}-\mathbf{q}/2\right)}$
and $\ket{\psi_{m}\left(\mathbf{k}+\mathbf{q}/2\right)}$, for which 
\begin{equation}
\begin{aligned}\bra{\psi_{n}\left(\mathbf{k}-\mathbf{q}/2\right)}\hat v_{a}\left(\mathbf{q}\right)\ket{\psi_{m}\left(\mathbf{k}+\mathbf{q}/2\right)} & =-ie\sum_{\Delta\mathbf{R}}e^{i\mathbf{k}\cdot\Delta\mathbf{R}}\Delta R^{a}h^{nm}_{\Delta\mathbf{R}}\\
 & =v_{a}\left(\mathbf{k}\right)
\end{aligned}
\label{eq:vk}
\end{equation}

Thus, within a Bloch basis, the real-space paramagnetic operator reduces
exactly to the conventional velocity operator between states separated
by momentum $\mathbf{q}$. The trace over the full Hilbert space appearing
in Eq.\eqref{eq:Reference_Parm_Sup} may now be expressed in the Bloch basis
as 
\begin{equation}
\text{Tr}\left[...\right]=\sum_{n}\int_{\text{FBZ}}d\mathbf{k}\bra{\psi_{n}\left(\mathbf{k}-\mathbf{q}/2\right)}\left[...\right]\ket{\psi_{n}\left(\mathbf{k}-\mathbf{q}/2\right)}
\end{equation}
Using $\hat G^{R}-\hat G^{A}=-2\pi i\delta(\varepsilon-\hat H)$, the paramagnetic contribution
becomes
\begin{equation}
\begin{alignedat}{1}\Pi_{ab}(\omega,\mathbf{q}) & =\frac{ie^2}{2\pi\hbar}\int d\varepsilon d\mathbf{k}f(\varepsilon)\left[\Delta \hat G_{nm}\left(\mathbf{k}-\frac{\mathbf{q}}{2},\varepsilon\right)\hat v^{md}_{a}\left(\mathbf{k},\mathbf{q}\right)\hat G^{R}_{dl}(\mathbf{k}+\frac{\mathbf{q}}{2},\varepsilon+\omega)\hat v^{ln}_{b}\left(\mathbf{k},-\mathbf{q}\right)\right.\\
 & \qquad\qquad\qquad\qquad\,\,\,+\left.\hat G^{A}_{nm}(\mathbf{k}-\frac{\mathbf{q}}{2},\varepsilon-\omega)\hat v^{md}_{a}\left(\mathbf{k},+\mathbf{q}\right)\Delta \hat G_{dl}\left(\mathbf{k}+\frac{\mathbf{q}}{2},\varepsilon\right)\hat v^{ln}_{b}\left(\mathbf{k},-\mathbf{q}\right)\right],
\end{alignedat}
\label{eq:intermediate_bloch}
\end{equation}
where 
\begin{equation}
\begin{aligned}\hat v^{nm}_{a}\left(\mathbf{k},\mathbf{q}\right) & =\bra{\psi_{n}\left(\mathbf{k}-\mathbf{q}/2\right)}\hat v_{a}\left(\mathbf{q}\right)\ket{\psi_{m}\left(\mathbf{k}+\mathbf{q}/2\right)}\end{aligned}
\end{equation}
and
\begin{equation}
\hat G^{R/A}_{mn}(\mathbf{k},\varepsilon)=\left(\frac{1}{\varepsilon-\hat H\left(\mathbf{k}\right)\pm i0^{+}}\right)_{mn}.
\end{equation}
Inserting Equation (\ref{eq:vk}) in (\ref{eq:intermediate_bloch})
one obtains 
\begin{equation}
\begin{alignedat}{1}\Pi_{ab}(\omega,\mathbf{q}) & =\frac{ie^2}{2\pi\hbar}\int d\varepsilon d\mathbf{k}f(\varepsilon)\left[\Delta \hat G_{nm}\left(\mathbf{k}-\frac{\mathbf{q}}{2},\varepsilon\right)\hat v^{md}_{a}\left(\mathbf{k}\right)\hat G^{R}_{dl}(\mathbf{k}+\frac{\mathbf{q}}{2},\varepsilon+\omega) \hat v^{ln}_{b}\left(\mathbf{k}\right)\right.\\
 & \qquad\qquad\qquad\qquad\,+\left.\hat G^{A}_{nm}(\mathbf{k}-\frac{\mathbf{q}}{2},\varepsilon-\omega)]\hat v^{md}_{a}\left(\mathbf{k}\right)\Delta \hat G_{dl}\left(\mathbf{k}+\frac{\mathbf{q}}{2},\varepsilon\right)\hat v^{ln}_{b}\left(\mathbf{k}\right)\right].
\end{alignedat}
\end{equation}
Finally, recognizing that the remaining indices refer exclusively
to the internal degrees of freedom within the unit cell, the expression
may be written compactly as a trace over the Bloch spinor space, 
\begin{equation}
\begin{alignedat}{1}\Pi_{ab}(\omega,\mathbf{q}) & =\frac{ie^2}{2\pi\hbar}\int d\varepsilon d\mathbf{k}f(\varepsilon)\text{Tr}_{\mathbf{k}}\left[\Delta \hat G\left(\mathbf{k}-\frac{\mathbf{q}}{2},\varepsilon\right)\hat v_{a}\left(\mathbf{k}\right)\hat G^{R}(\mathbf{k}+\frac{\mathbf{q}}{2},\varepsilon+\omega)\hat v_{b}\left(\mathbf{k}\right)\right.\\
 & \qquad\qquad\qquad\qquad\qquad+\left.\hat G^{A}(\mathbf{k}-\frac{\mathbf{q}}{2},\varepsilon-\omega)\hat v_{a}\left(\mathbf{k}\right)\Delta \hat G\left(\mathbf{k}+\frac{\mathbf{q}}{2},\varepsilon\right)\hat v_{b}\left(\mathbf{k}\right)\right],
\end{alignedat}
\end{equation}
which is precisely the paramagnetic contribution derived by$\,$\citep{shinadaOrbitalOpticalActivity2024}
for translationally invariant systems. Thus, the real-space formulation
developed in the present work is fully consistent with the conventional
momentum-space theory in the crystalline limit.

\section{Vanishing of Static Contribution} \label{Ap:Static_Contribution}

Following a thought process
close to that of Ref. \citep{shinadaOrbitalOpticalActivity2024},
we now prove that the static contribution $\Phi_{ab,c}\left(0\right)$
is totally antisymmetric and therefore vanishes for two dimensional
systems, irrespectively of the vector nature of $\mathbf{E}$. 

The zeroth order term in Eq. (\ref{eq:phi_omega_expansion}) is written
in as

\begin{equation}
\begin{aligned}
\begin{aligned}
\Phi_{ab,c}\left(0\right)=
\frac{ie^2}{2\hbar}\int d\varepsilon f\left(\varepsilon\right)\text{Tr}\left[\delta\left(\varepsilon-\hat H\right)\left\{ \hat r_{c},\hat v_{a}\right\} \hat G^{R}(\varepsilon)\hat v_{b}-\delta\left(\varepsilon-\hat H\right)\hat v_{a} \hat G^{R}(\varepsilon)\left\{ \hat r_{c},\hat v_{b}\right\} \right.\\
\left.-\delta\left(\varepsilon-\hat H\right)\left\{ \hat r_{c},\hat v_{b}\right\} \hat G^{A}(\varepsilon)\hat v_{a}+\delta\left(\varepsilon-\hat H\right)\hat v_{b}\hat G^{A}(\varepsilon)\left\{ \hat r_{c},\hat v_{a}\right\} \right].
\end{aligned}
\end{aligned}
\end{equation}

We can use $-2\pi i\delta\left(\varepsilon-\hat H\right)= \hat G^{R}\left(\varepsilon\right)-\hat G^{A}\left(\varepsilon\right)$
to write 
\begin{equation}
\begin{aligned}
\Phi_{ab,c}\left(0\right) & =
\frac{ie^2}{2\hbar}\int d\varepsilon f\left(\varepsilon\right)\text{Tr}\left[\left[\delta\left(\varepsilon-\hat H\right),\hat r_{c}\right]\hat v_{a}\hat G^{R}(\varepsilon)\hat v_{b}+\delta\left(\varepsilon-\hat H\right)\hat v_{a}\left[\hat r_{c},\hat G^{R}(\varepsilon)\right]\hat v_{b}\right.\\
 & \qquad\qquad\qquad\qquad-\!\!\left.\left[\delta\left(\varepsilon-\hat H\right),\hat r_{c}\right]\hat v_{b}\hat G^{A}(\varepsilon)\hat v_{a}-\delta\left(\varepsilon-H\right)v_{b}\left[R_{c},G^{A}(\varepsilon)\right]v_{a}\right],
\end{aligned}
\end{equation}
This expression is equivalent to
\begin{equation}
\begin{aligned}
\Phi_{ab,c}\left(0\right) & =-
\frac{e^2}{4\pi\hbar}\int d\varepsilon f\left(\varepsilon\right)\text{Tr}\left[\left[\Delta \hat G\left(\varepsilon\right),\hat r_{c}\right]\hat v_{a}\hat G^{R}(\varepsilon)\hat v_{b}+\Delta \hat G\left(\varepsilon\right)\hat v_{a}\left[\hat r_{c},\hat G^{R}(\varepsilon)\right]\hat v_{b}\right]+\\
 & \phantom{{}=-\frac{1}{4\pi\hbar}\int d\varepsilon f\left(\varepsilon\right)} \text{Tr}\left[\left[\Delta \hat G\left(\varepsilon\right),\hat r_{c}\right]\hat v_{b}\hat G^{A}(\varepsilon)\hat v_{a}+\Delta \hat G\left(\varepsilon\right)\hat v_{b}\left[\hat r_{c},\hat G^{A}(\varepsilon)\right]\hat v_{a}\right], 
\end{aligned}
\end{equation}
where $G^{R}\left(\varepsilon\right)-\hat G^{A}\left(\varepsilon\right) = \Delta \hat G\left(\varepsilon\right)$. Denoting $\left[G^{R/A}\left(\varepsilon\right)\right]_{-}\equiv\left[\hat r_{c},G^{R/A}\left(\varepsilon\right)\right]$ and using the cyclic property of the trace we are left with
\begin{equation}
\begin{aligned}\begin{aligned}\Phi_{ab,c}\left(0\right) & =-
\frac{e^2}{4 \pi\hbar}\int d\varepsilon f\left(\varepsilon\right)\text{Tr}\left[-\left[\hat G^{R}\right]_{-}\hat v_{a}\hat G^{R}\hat v_{b}+G^{R}v_{a}\left[\hat G^{R}\right]_{-}\hat v_{b}+\left[\hat G^{A}\right]_{-}\hat v_{b}\hat G^{A}\hat v_{a}-\hat G^{A}\hat v_{b}\left[\hat G^{A}\right]_{-}\hat v_{a}\right],
\end{aligned}
\end{aligned}
\label{eq:phi_ijc_static_comms-1}
\end{equation}

where the energy arguments of the Green's functions where suppressed for brevity. 

Since $\left[G^{R/A}\left(\varepsilon\right)\right]^{*}_{-}=-\left[G^{A/R}\left(\varepsilon\right)\right]_{-}$,
we can see that
\begin{equation}
\text{Tr}\left[\left[\hat G^{A}\left(\varepsilon\right)\right]_{-}\hat v_{b}\hat G^{A}\left(\varepsilon\right)\hat v_{a}\right]=-\text{Tr}\left[\hat G^{R}\left(\varepsilon\right) \hat v_{b}\left[\hat G^{R}\left(\varepsilon\right)\right]_{-}\hat v_{a}\right]^{*},
\end{equation}
and therefore write Eq.\eqref{eq:phi_ijc_static_comms-1} as
\begin{equation}
\begin{aligned}\begin{aligned}\Phi_{ab,c}\left(0\right) & =-\frac{e^2}{2\pi\hbar}\text{Re}\int d\varepsilon f\left(\varepsilon\right)\text{Tr}\left[-\left[\hat r_{c},\hat G^{R}\left(\varepsilon\right)\right]\hat v_{a}\hat G^{R}\left(\varepsilon\right)\hat v_{b}+\hat G^{R}\left(\varepsilon\right)\hat v_{a}\left[\hat r_{c},\hat G^{R}\left(\varepsilon\right)\right]\hat v_{b}\right].
\end{aligned}
\end{aligned}
\end{equation}
Now we can use the identity 
\begin{equation}
\left[\hat r_{c},\hat G^{R/A}(\varepsilon)\right]=-\hat G^{R/A}(\varepsilon)\left[\hat H,\hat r_{c}\right]\hat G^{R/A}(\varepsilon)=i\hbar\,\hat G^{R/A}(\varepsilon)\hat v_{c}\hat G^{R/A}(\varepsilon), \label{eq:green_position_velocity}
\end{equation}
to write
\begin{equation}
\begin{aligned}\Phi_{ab,c}\left(0\right) & =\frac{e^2}{2\pi \hbar}\text{Im}\int d\varepsilon f\left(\varepsilon\right)\text{Tr}\left[\hat G^{R}\left(\varepsilon\right)\hat v_{c}\hat G^{R}\left(\varepsilon\right)\hat v_{a}\hat G^{R}\left(\varepsilon\right) \hat v_{b}-\hat G^{R}\left(\varepsilon\right)\hat v_{a}\hat G^{R}\left(\varepsilon\right)\hat v_{c}\hat G^{R}\left(\varepsilon\right) \hat v_{b}\right]\end{aligned}
.\label{eq:phi_static_antisymmetric-1}
\end{equation}

It is evident that Eq. (\ref{eq:phi_static_antisymmetric-1}) is completely
antisymmetric under the exchange of any two indices $a,b$ and $c$.
It can be written as 
\begin{equation}
\Phi_{ab,c}(0)=\epsilon_{abc}\times \text{const},\label{eq:static_antisymmetric}
\end{equation}
which implies that the static limit for 2D systems vanishes both for $\mathbf{E}\parallel \mathbf{q}$ and $\mathbf{E}\perp \mathbf{q}$. Since the calculations presented in the main text concern systems subject to longitudinal electric fields (dc transport limit), one has $\Phi_{ab,c}(0)=0$ and the dc response is entirely determined by the first frequency derivative of the response kernel.

\section{Středa Decomposition of the Linear-in-q Conductivity} 
\label{Ap:Streda_Decomposition}

Following the symmetry-based decomposition proposed in Ref. $\,$\citep{bonbien_symmetrized_2020},
we separate the conductivity according to the exchange symmetry of
the two current operators appearing in the Kubo formula. This operation
is equivalent to simultaneously exchanging the Cartesian indices and
reversing the transferred momentum, 
\begin{equation}
\sigma_{ab}\left(\omega,\mathbf{q}\right)=\underbrace{\frac{1}{2}\left(\sigma_{ab}\left(\omega,\mathbf{q}\right)+\sigma_{ba}\left(\omega,-\mathbf{q}\right)\right)}_{\sigma^{\text{A}}_{ab}\left(\omega,\mathbf{q}\right)}+\underbrace{\frac{1}{2}\left(\sigma_{ab}\left(\omega,\mathbf{q}\right)-\sigma_{ba}\left(\omega,-\mathbf{q}\right)\right).}_{\sigma^{\text{S}}_{ab}\left(\omega,\mathbf{q}\right)}\label{eq:cond_separate}
\end{equation}
The physical interpretation of this decomposition follows directly
from the Onsager reciprocity relations,

\begin{equation}
\sigma_{ab}\left(\omega,\mathbf{q},\mathbf{B}\right)=\sigma_{ba}\left(\omega,-\mathbf{q},-\mathbf{B}\right).
\end{equation}
For time-reversal-symmetric systems ($\mathbf{B}=0$), the conductivity
satisfies 
\begin{equation}
\sigma_{ab}\left(\omega,\mathbf{q}\right)=\sigma_{ba}\left(\omega,-\mathbf{q}\right),
\end{equation}
so that the second contribution in Eq. (\ref{eq:cond_separate}) vanishes
identically. Breaking time-reversal symmetry, on the other hand, allows
this second contribution to become finite. Accordingly, we refer to
the two sectors $\sigma^{\text{A}}_{ab}\left(\omega,\mathbf{q}\right)$
and $\sigma^{\text{S}}_{ab}\left(\omega,\mathbf{q}\right)$ as the
time-even (Fermi-surface-like) and time-odd (Fermi-sea-like) contributions,
respectively. The reason for the superscripts A and S will become clear once we linearize in $\mathbf q$. 

Expanding the decomposition in Eq. (\ref{eq:cond_separate}) to first
order in q yields 
\begin{equation}
\begin{alignedat}{1}\sigma^{\text{A}}_{ab}\left(\omega,\mathbf{q}\right) & =\sigma^{\text{A}}_{ab}\left(\omega\right)+\sigma^{\text{A}}_{ab,c}\left(\omega\right)q_{c}+...\\
\sigma^{\text{S}}_{ab}\left(\omega,\mathbf{q}\right) & =\sigma^{\text{S}}_{ab}\left(\omega\right)+\sigma^{\text{S}}_{ab,c}\left(\omega\right)q_{c}+...
\end{alignedat}
\end{equation}
As before, $\sigma^{\text{A}/\text{S}}_{ab}\left(\omega\right)$
denotes the decomposition of the response to a spatially uniform electric
field, whereas $\sigma^{\text{A}/\text{S}}_{ab,c}\left(\omega\right)$
corresponds to the linear correction in the transferred momentum.
The latter are given by 
\begin{equation}
\begin{aligned}\sigma^{\text{A}}_{ab,c}\left(\omega\right) & =\frac{1}{2}\left(\sigma_{ab,c}\left(\omega\right)-\sigma_{ba,c}\left(\omega\right)\right),\\
\sigma^{\text{S}}_{ab,c}\left(\omega\right) & =\frac{1}{2}\left(\sigma_{ab,c}\left(\omega\right)+\sigma_{ba,c}\left(\omega\right)\right).
\end{aligned}
\label{eq:linear_conds}
\end{equation}
It is now clear why we indexed the time-even and the time-odd sectors by A and S respectively. By construction, $\sigma^{\text{A}}_{ab,c}\left(\omega\right)$
is antisymmetric under the exchange of the first two Cartesian indices,
whereas $\sigma^{\text{S}}_{ab,c}\left(\omega\right)$ is symmetric. 

\subsection{Time-even contribution}
To identify the dc limit of the time-even response, we substitute
Eq.$\,$\eqref{eq:derivative_phi} in Eq.$\,$\eqref{eq:linear_conds}
and take the limit $\omega\rightarrow0$, obtaining 
\begin{equation}
\begin{aligned}\sigma^{\text{A}}_{ab,c} & =\frac{e^2}{4\hbar}\int d\varepsilon\,f(\varepsilon)\,\mathrm{Tr}\left[\delta(\varepsilon-\hat H)\{\hat r_{c},\hat v_{a}\}\partial_{\varepsilon}\hat G^{R}(\varepsilon)\hat v_{b}-\delta(\varepsilon-\hat H)v_{a}\partial_{\varepsilon}\hat G^{R}(\varepsilon)\{\hat r_{c},\hat v_{b}\}\right]\\
 & \qquad\qquad\qquad\qquad-\,\mathrm{Tr}\left[\delta(\varepsilon-\hat H)\hat v_{b}\partial_{\varepsilon}\hat G^{A}(\varepsilon)\{\hat r_{c},\hat v_{a}\}-\delta(\varepsilon-\hat H)\{\hat r_{c},\hat v_{b}\}\partial_{\varepsilon}\hat G^{A}(\varepsilon)\hat v_{a}\right]\\
 & \qquad\qquad\qquad\qquad-\,\mathrm{Tr}\left[\delta(\varepsilon-\hat H)\{\hat r_{c},\hat v_{b}\}\partial_{\varepsilon}\hat G^{R}(\varepsilon)\hat v_{a}-\delta(\varepsilon-\hat H)\hat v_{b}\partial_{\varepsilon}\hat G^{R}(\varepsilon)\{\hat r_{c},\hat v_{a}\}\right]\\
 & \qquad\qquad\qquad\qquad+\,\mathrm{Tr}\left[\delta(\varepsilon-\hat H)\hat v_{a}\partial_{\varepsilon}\hat G^{A}(\varepsilon)\{\hat r_{c},\hat v_{b}\}-\delta(\varepsilon-\hat H)\{\hat r_{c},\hat v_{a}\}\partial_{\varepsilon}\hat G^{A}(\varepsilon)\hat v_{b}\right],
\end{aligned}
\end{equation}
or equivalently
\begin{equation}
\begin{aligned}\sigma^{\text{A}}_{ab,c} & =\frac{ie^2}{8\pi\hbar}\int d\varepsilon\,f(\varepsilon)\,\mathrm{Tr}\left[\Delta \hat G\left(\varepsilon\right) \{\hat r_{c},\hat v_{a}\}\partial_{\varepsilon}\Delta \hat G \left(\varepsilon\right) \hat v_{b}-\Delta \hat G \left(\varepsilon\right) \hat v_{a}\partial_{\varepsilon}\Delta \hat G \left(\varepsilon\right)\{\hat r_{c},\hat v_{b}\}\right]\\
 & \qquad\qquad\qquad\qquad+\,\mathrm{Tr}\left[\Delta \hat G \left(\varepsilon\right) \hat v_{b}\partial_{\varepsilon}\Delta \hat G \left(\varepsilon\right) \{\hat r_{c},\hat v_{a}\}-\Delta \hat G \left(\varepsilon\right) \{\hat r_{c},\hat v_{b}\}\partial_{\varepsilon}\Delta \hat G \left(\varepsilon\right) \hat v_{a}\right].
\end{aligned}
\end{equation}
Since the boundary term vanishes, integrating by parts brings out
the Fermi-surface term 
\begin{equation}
\begin{aligned}\sigma^{\text{A}}_{ab,c} & =-\frac{ie^2}{8\pi\hbar}\int d\varepsilon\,\,\mathrm{Tr}\left[\partial_{\varepsilon}\left(f(\varepsilon)\Delta \hat G \left(\varepsilon\right)\right)\{\hat r_{c},\hat v_{a}\}\Delta \hat G \left(\varepsilon\right) \hat v_{b}-\partial_{\varepsilon}\left(f(\varepsilon)\Delta \hat G \left(\varepsilon\right) \right)\hat v_{a}\Delta \hat G \left(\varepsilon\right) \{\hat r_{c},\hat v_{b}\}\right]\\
 &\!\!\!\!\quad\qquad\qquad+\,\mathrm{Tr}\left[\partial_{\varepsilon}\left(f(\varepsilon)\Delta \hat G \left(\varepsilon\right) \right)\hat v_{b}\Delta \hat G\left(\varepsilon\right)\{\hat r_{c}, \hat v_{a}\}-\partial_{\varepsilon}\left(f(\varepsilon)\Delta \hat G\left(\varepsilon\right) \right)\{\hat r_{c},\hat v_{b}\}\Delta \hat G \left(\varepsilon\right) \hat v_{a}\right].
\end{aligned}
\end{equation}
The derivative now acts on the product $f\left(\varepsilon\right)\Delta \hat G$,
naturally separating the response into two contributions, 
\begin{equation}
\sigma^{\text{A}}_{ab,c}=\sigma^{\text{FS}}_{ab,c}+\sigma^{\Delta}_{ab,c},\label{eq:fsurf_separation}
\end{equation}
where $\sigma^{\text{FS}}_{ab,c}$ is proportional to the derivative
of the Fermi-Dirac distribution, representing a genuine Fermi-surface
contribution, while $\sigma^{\Delta}_{ab,c}$ originates from the
energy derivative of the spectral function. These two terms are given
by 
\begin{equation}
\begin{aligned}\sigma^{\text{FS}}_{ab,c} & =-\frac{ie^2}{8\pi\hbar}\int d\varepsilon\,\partial_{\varepsilon}f(\varepsilon)\,\mathrm{Tr}\left[\Delta \hat G \left(\varepsilon\right)\{\hat r_{c},\hat v_{a}\}\Delta \hat G \left(\varepsilon\right) \hat v_{b}-\Delta \hat G\left(\varepsilon\right) \hat v_{a}\Delta \hat G \left(\varepsilon\right) \{\hat r_{c},\hat v_{b}\}\right]\\
 & \qquad\qquad\qquad\qquad+\,\mathrm{Tr}\left[\Delta \hat G\left(\varepsilon\right) \hat v_{b}\Delta \hat G\left(\varepsilon\right)\{\hat r_{c},\hat v_{a}\}-\Delta \hat G\left(\varepsilon\right)\{\hat r_{c},\hat v_{b}\}\Delta \hat G \left(\varepsilon\right) \hat v_{a}\right].\\
 & =-\frac{ie^2}{4\pi\hbar}\int d\varepsilon\,\partial_{\varepsilon}f(\varepsilon)\,\mathrm{Tr}\left[\Delta \hat G \left(\varepsilon\right) \{\hat r_{c},\hat v_{a}\}\Delta \hat G \left(\varepsilon\right) \hat v_{b}-\Delta \hat G \left(\varepsilon\right) \hat v_{a}\Delta \hat G\left(\varepsilon\right) \{\hat r_{c},\hat v_{b}\}\right]
\end{aligned}
\end{equation}
and 
\begin{equation}
\begin{aligned}\sigma^{\Delta}_{ab,c} & =-\frac{ie^2}{8\pi\hbar}\int d\varepsilon\,f(\varepsilon)\,\mathrm{Tr}\left[\partial_{\varepsilon}\Delta \hat G \left(\varepsilon\right) \{\hat r_{c},\hat v_{a}\}\Delta \hat G \hat v_{b}-\partial_{\varepsilon}\Delta \hat G \left(\varepsilon\right) \hat v_{a}\Delta \hat G \left(\varepsilon\right) \{\hat r_{c},\hat v_{b}\}\right]\\
 & \,\,\,\qquad\qquad\qquad+\,\mathrm{Tr}\left[\partial_{\varepsilon}\Delta \hat G \left(\varepsilon\right) \hat v_{b}\Delta \hat G\left(\varepsilon\right)\{\hat r_{c},\hat v_{a}\}-\partial_{\varepsilon}\Delta \hat G \left(\varepsilon\right)\{\hat r_{c},\hat v_{b}\}\Delta \hat G\left(\varepsilon\right) \hat v_{a}\right],
\end{aligned}
\label{eq:sigma_RA}
\end{equation}
Using the cyclic property of the trace on Eq.$\,$\eqref{eq:sigma_RA},
one finds 
\begin{equation}
\sigma^{\Delta}_{ab,c}=-\sigma^{\text{A}}_{ab,c}.\label{eq:RA_fsurf}
\end{equation}
which immediately yields the final expression for the time-even contribution,
\begin{equation}
\begin{aligned}\sigma^{\text{A}}_{ab,c} & =-\frac{ie^2}{8\pi\hbar}\int d\varepsilon\,\partial_{\varepsilon}f(\varepsilon)\,\mathrm{Tr}\left[\Delta \hat G\left(\varepsilon\right)\{\hat r_{c},\hat v_{a}\}\Delta \hat G\left(\varepsilon\right) \hat v_{b}-\Delta \hat G\left(\varepsilon\right) \{\hat r_{c},\hat v_{b}\}\Delta \hat G\left(\varepsilon\right) \hat v_{a}\right]\end{aligned}
.\label{eq:fsurf}
\end{equation}

Equation$\,$\eqref{eq:fsurf} is manifestly antisymmetric under the
exchange $a\leftrightarrow b$, as required by the definition of the
time-even response. Moreover, because it depends exclusively on the
spectral function $\Delta \hat G\left(\varepsilon\right)$, it is entirely determined by states
at the Fermi level and therefore represents a purely Fermi-surface
contribution. This is consistent with previous results obtained for
translationally invariant systems in the dc limit \citep{pozo_ocana_multipole_2023}.  

\subsection{Time-odd contribution}
Owing to its symmetry under the exchange $a\leftrightarrow b$, the
time-odd contribution admits a considerably simpler derivation. Following
the same procedure as for the time-even sector, one obtains 
\begin{equation}
\begin{aligned}\sigma^{\text{S}}_{ab,c} & =\frac{ie^2}{8\pi\hbar}\int d\varepsilon\,f(\varepsilon)\,\mathrm{Tr}\left[\Delta \hat G \left(\varepsilon\right) \{\hat r_{c},\hat v_{a}\}\partial_{\varepsilon}\left(G^{R}\left(\varepsilon\right)+G^{A}\left(\varepsilon\right)\right)\hat v_{b}-\Delta \hat G\left(\varepsilon\right) \hat v_{a}\partial_{\varepsilon}\left(G^{R}\left(\varepsilon\right)+G^{A}\left(\varepsilon\right)\right)\{\hat r_{c},\hat v_{b}\}\right]\\
 & \qquad\qquad\qquad+\mathrm{Tr}\left[\Delta \hat G\left(\varepsilon\right) \hat v_{b}\partial_{\varepsilon}\left(\hat G^{R}\left(\varepsilon\right)+\hat G^{A}\left(\varepsilon\right)\right)\{\hat r_{c},\hat v_{a}\}-\Delta \hat G\left(\varepsilon\right)\{\hat r_{c},\hat v_{b}\}\partial_{\varepsilon}\left(\hat G^{R}\left(\varepsilon\right)+\hat G^{A}\left(\varepsilon\right)\right)\hat v_{a}\right].
\end{aligned}
\end{equation}
Using the relation
\begin{equation}
    \text{Tr}\left[\Delta \hat G\left(\varepsilon\right) \hat v_{b}\partial_{\varepsilon}\left(\hat G^{R}+\hat G^{A}\right)\{\hat r_{c},\hat v_{a}\}\right]^{*}=\text{Tr}\left[\Delta \hat G\left(\varepsilon\right)\{\hat r_{c},\hat v_{a}\}\partial_{\varepsilon}\left(\hat G^{R}\left(\varepsilon\right)+\hat G^{A}\left(\varepsilon\right)\right)\hat v_{b}\right],
\end{equation}
the response can be written in the compact form

\begin{align}\sigma^{\text{S}}_{ab,c} & =\frac{ie^2}{4\pi\hbar}\text{Re}\int d\varepsilon\,f(\varepsilon)\,\mathrm{Tr}\left[\Delta \hat G\left(\varepsilon\right)\{\hat r_{c},\hat v_{a}\}\partial_{\varepsilon}\left(\hat G^{R}\left(\varepsilon\right)+\hat G^{A}\left(\varepsilon\right)\right)\hat v_{b}+\Delta \hat G\left(\varepsilon\right)\{\hat r_{c},\hat v_{b}\}\partial_{\varepsilon}\left(\hat G^{R}\left(\varepsilon\right)+\hat G^{A}\left(\varepsilon\right)\right)\hat v_{a}\right].
\end{align}

This expression is manifestly symmetric under the exchange $a\leftrightarrow b$,
in accordance to its definition. We note that in weakly disordered systems,
the Fermi-sea-like contribution to the uniform Hall conductivity has been traditionally identified with the intrinsic
response (though the decomposition  of linear-response quantities into Fermi-surface and Fermi-sea-like contributions is not unique \cite{Joao_2025}). For the linear-in-$\mathbf{q}$ conductivity, however, this identification
is generally not possible. As shown previously for Bloch electrons \citep{pozo_ocana_multipole_2023},
the symmetric component contains a Fermi-surface term
that survives in the dc limit. This behavior suggests that $\sigma^\text{S}_{abc}$ receives significant extrinsic contributions even in the weak disorder limit, as indeed our current work demonstrates.

\end{document}